\documentclass[conference]{IEEEtran}
\makeatletter

\def\ps@headings{%
	
	\def\@oddhead{\mbox{}\scriptsize\rightmark \hfil \thepage}%
	
	\def\@evenhead{\scriptsize\thepage \hfil \leftmark\mbox{}}%
	
	\def\@oddfoot{}%
	
	\def\@evenfoot{}}

\makeatother

\IEEEoverridecommandlockouts
\usepackage{cite}
\usepackage{amsmath,amssymb,amsfonts}
\usepackage{algorithmic}
\usepackage{graphicx}
\usepackage{textcomp}
\usepackage{xcolor}
\def\BibTeX{{\rm B\kern-.05em{\sc i\kern-.025em b}\kern-.08em
		T\kern-.1667em\lower.7ex\hbox{E}\kern-.125emX}}

\usepackage{caption}
\usepackage{graphicx,subfigure}
\usepackage{float} 
\usepackage{bm}
\usepackage{enumerate}
\usepackage{color}
\usepackage{amsfonts}
\usepackage{amsmath}
\usepackage{mathrsfs}
\usepackage{algorithmic}
\usepackage{algorithm}
\usepackage{overpic}
\usepackage{epsfig}
\usepackage{tensor}
\usepackage{latexsym}
\usepackage{amssymb}
\usepackage{amsmath}
\usepackage{graphicx}
\usepackage{subfigure}
\usepackage{url}

\usepackage{caption}
\usepackage{multirow}
\usepackage{epstopdf}
\usepackage{diagbox}
\usepackage{enumerate}

\usepackage{caption}
\usepackage{algorithm}
\usepackage{algorithmic}
\usepackage{graphicx}
\usepackage{enumerate}
\usepackage{amsthm}
\usepackage{epstopdf}
\usepackage{amsmath,amssymb,epsfig,graphicx,slidesec,colortbl}
\usepackage[absolute,overlay]{textpos}
\usepackage{multimedia,algorithm,algorithmic,multirow,url}
\usepackage{cite}
\usepackage{subfig}
\usepackage{color}
\usepackage{enumitem}

\usepackage{amsmath} 
\usepackage{enumitem}  
\usepackage{indentfirst} 
\usepackage{stfloats}
\usepackage{enumerate}
\newtheorem{observation}{Observation}
\newtheorem{defin}{Definition}
\newtheorem{prop}{Proposition}
\newtheorem{cor}{Corollary}
\newtheorem{lem}{Lemma}
\newtheorem{gam}{Game}

\newtheorem{thm}{Theorem}

\begin{document}
	
	\title{An Incentive Mechanism for Sustainable Blockchain Storage}
	
	%
	
	\author{
		
		\IEEEauthorblockN{Yunshu Liu, \emph{Student Member, IEEE}, Zhixuan Fang, \emph{Member, IEEE}, Man Hon Cheung,\\ Wei Cai, \emph{Member, IEEE}, and Jianwei Huang, \emph{Fellow, IEEE}}
		
		%
		%
		%
		\thanks{Part of this paper was presented in ICC 2020 \cite{liueconomics}. (Corresponding author: Jianwei Huang.)
			
			Y. Liu is with the Department of Information Engineering, The Chinese University of Hong Kong, Hong Kong and Shenzhen Institute of Artificial Intelligence and Robotics for Society (AIRS), Shenzhen, China. 
			
			Z. Fang is with Institute for Interdisciplinary Information Sciences, Tsinghua University, Beijing, China; and Shanghai Qi Zhi Institute, Shanghai, China. 
			
			M. H. Cheung is with the Department of Computer Science, City University of Hong Kong, Hong Kong. 
			
			W. Cai is with the School of Science and Engineering, The Chinese University of Hong Kong, Shenzhen, China, and Shenzhen Institute of Artificial Intelligence and Robotics for Society (AIRS), Shenzhen, China.	
			
			J. Huang is with the School of Science and Engineering, The Chinese University of Hong Kong, Shenzhen, China, and Shenzhen Institute of Artificial Intelligence and Robotics for Society (AIRS), Shenzhen, China (corresponding author).
		}
	}
	
	
	\maketitle
	
	
	\begin{abstract}
		Miners in a blockchain system are suffering from ever-increasing storage costs, which in general have not been properly compensated by the users' transaction fees. This reduces the incentives for the miners' participation and may jeopardize the blockchain security. We propose to mitigate this blockchain insufficient fee issue through a Fee and Waiting Tax (FWT) mechanism, which explicitly considers the two types of negative externalities in the system. 
		Specifically, we model the interactions between the protocol designer, users, and miners as a three-stage Stackelberg game. 
		By characterizing the equilibrium of the game, we find that miners neglecting the negative externality in transaction selection cause they are willing to accept insufficient-fee transactions. This leads to the insufficient storage fee issue in the existing protocol. Moreover, our proposed optimal FWT mechanism can motivate users to pay sufficient transaction fees to cover the storage costs and achieve the unconstrained social optimum. 
		Numerical results show that the optimal FWT mechanism guarantees sufficient transaction fees and achieves an average social welfare improvement of 33.73\% or more over the existing protocol. Furthermore, the optimal FWT mechanism achieves the maximum fairness index, and performs well even under heterogeneous-storage-cost miners.
		
	\end{abstract}
	\maketitle
	
	\section*{Nomenclature}
	\begin{table}[h]
		\normalsize 
		\begin{tabular}{ll}
			\textit{\fontsize{13}{10}Sets} & \\
			$\mathcal{N}$ & Set of users\\
			$\mathcal{N}_H(\mathcal{N}_L)$ & Set of type-$H$ (type-$L$) users\\
			\vspace{3mm}$\mathcal{M}$ & Set of miners\\
			\textit{\fontsize{13}{10}Indices}& \\
			$n$ & Index of users\\
			$m$ & Index of miners\\
		\end{tabular}
	\end{table}
	\begin{table}[h]
		\normalsize 
		\begin{tabular}{ll}
			$(n,i)$ & Index of $i$-th transaction of user $n$\\
			\vspace{3mm}$k$ & Index of rounds of mining\\
			
			\textit{\fontsize{13}{10}Variables}& \\
			$\boldsymbol{\rho}$ & Fee-per-byte choice\\
			$\overline{\rho}$, $\underline{\rho}$ & Two fee-per-byte choices\\
			$\boldsymbol{P}$ & Waiting tax rate vector\\
			$P_{HH} (P_{HL})$ & Waiting tax that a type-$H$ user pays to \\
			& another type-$H$ (type-$L$) user\\
			$P_{LH} (P_{LL})$ & Waiting tax that a type-$L$ user pays to \\
			&another type-$H$ (type-$L$) user\\
			$p_{nl}$ & Waiting tax that user $n$ pays to user $l$\\
			$\boldsymbol{\lambda}$ & All users' transaction generation rates\\
			$\boldsymbol{\lambda}_{n}$ & User $n$'s transaction generation rates\\
			$\lambda_{n1},\lambda_{n2}$ & User $n$'s transaction generation rates at $\overline{\rho}$ and $\underline{\rho}$\\
			\vspace{3mm}$\mathcal{X}_m^k$ & Miner $m$'s transaction selection in round $k$\\
			
			\textit{\fontsize{13}{10}Parameters}& \\
			$N$ & Number of users\\
			$N_H(N_L)$ & Number of type-$H$ (type-$L$) users\\
			$M$ & Number of miners\\
			$\rho_{n,i}$ & Fee-per-byte of transaction $tx_{n,i}$\\
			$s_{n,i}$ & Size of transaction $tx_{n,i}$\\
			$\bar{s}$ & Expected size of transaction $tx_{n,i}$\\
			$w_{n,i}$ & Waiting time of transaction $tx_{n,i}$\\
			$R_n$ & User $n$'s transaction on-chain utility\\
			$R_H (R_L)$ & A type-$H$ (type-$L$) user's transaction on-chain\\
			&utility\\
			$\gamma$ & User's impatience level\\
			$C_s$ & A miner's storage cost per byte \\
			$\alpha_m$ & Miner $m$'s mining power\\
			$\mu$ & Block generation rate\\
			\vspace{3mm}$\mathcal{Q}^k$ & Transaction pool in round $k$\\

			\textit{\fontsize{13}{10}Payoffs}& \\
			$\theta_{n,i}$ & User $n$'s surplus from $tx_{n,i}$ \\
			$u_n$ & User $n$'s time-average payoff\\
			$v_m^k$ & Miner $m$'s payoff in round $k$\\
			$sw$ & Social welfare\\
		\end{tabular}
	\end{table}

	
	\section{Introduction}
	
	With the booming of cryptocurrencies, its underlying blockchain protocol imposes ever-increasing and significant storage costs on the solid-state storage drives \cite{Archive_node} of the operation nodes (often referred to as \emph{miners} \cite{Blockchain_survey}). For example, consider the second-largest cryptocurrency, Ethereum. Its data size grows by nearly 16 folds from 385 gigabytes in Feb. 2017 to 6.0 terabytes in Jan. 2021 \cite{Ethsize}. Currently, it costs each miner \$2000 per month to store the entire blockchain \cite{Archive_node}.	
	
	On the other hand, the blockchain users' transaction fee payments to the miners are insufficient to compensate such fast growth and significant storage costs. For example, the Ethereum users paid an average monthly transaction fee of \$7.32 million during the first half of 2020 \cite{Eth_fee}, much smaller than \$20 million monthly costs of all Ethereum nodes storing the entire blockchain (roughly 10,000 nodes \cite{No_full_node} and \$2000 monthly cost per node). The gap between the storage cost and transaction fee is filled by block reward, which is designed to gradually decrease over time in many blockchain systems (e.g., Bitcoin \cite{Bitcoin_white_paper}).
	
	With insufficient transaction fees, miners will have less incentive to stay in the system, which jeopardizes the blockchain system security. For example, the number of miners storing the Ethereum blockchain has declined 66\% since 2018, where the large storage costs could have played a major role \cite{Nodedecline}. The decline of miners may be catastrophic to the blockchain in the long run, as a less decentralized blockchain will become easier for the malicious miners to launch attacks\cite{BitcoinWiki_Weaknesses} and more vulnerable to a single point of failure \cite{cai2018decentralized}. To maintain a healthy decentralized ecosystem, it is critically important to mitigate the issue of insufficient transaction fees (for covering the storage costs).

	Mitigating the insufficient fee issue requires users to pay sufficient transaction fees to the miners. The protocol designer of the blockchain (e.g., the technical community serves as a leading role in protocol design) needs a proper mechanism to motivate this. However, to the best of our knowledge, there lacks enough \emph{theoretical} mechanism design work aiming at mitigating the insufficient fee issue (although the discussion in the technical community is heated \cite{Storage_discuss}). This motivates us to take the first step in this paper to propose such a mechanism to address the issue. 
	
	In this work, we focus on understanding the following two key questions: 
	
	%
	
	\begin{itemize}
		\item (i) \emph{Why are miners willing to accept insufficient-fee transactions in the existing blockchain system?}
		\item (ii) \emph{How to design an incentive mechanism to encourage users to pay sufficient transaction fees to the miners?}
	\end{itemize}
	
	To answer the above two questions, we propose a three-stage model to characterize the blockchain system. In Stage III, miners select transactions to include in the blockchain, considering the tradeoff between transaction fees and storage costs. In Stage II, users determine the transaction generation rates for different fees, by considering the tradeoff between paying high transaction fees and bearing high transaction waiting time. The transaction waiting time depends on how miners select transactions. In user-miner fee interaction of Stages II and III, there exist two types of negative externalities, i.e., each miner's transaction selection imposes storage costs on other miners and each user's transaction generation increases the average waiting time of other users. These two types of
	negative externalities are the reasons behind miners accepting insufficient-fee transactions and users experiencing excessive waiting time, respectively, in the existing protocol. Motivated by such an observation, we propose a Fee and Waiting Tax (FWT) mechanism for the protocol designer in Stage I. The mechanism determines the transaction fee choices for the users and meanwhile imposes waiting tax on the users, in order to motivate users to pay sufficient fees while achieving social welfare maximization.
	
	
	
	Our key results and contributions are as follows:\vspace{-2pt}

	
	\begin{itemize} [leftmargin=12pt]

		\item \textit{Fee mechanism design on blockchain storage:} To the best of our knowledge, this is one of the first theoretical studies on the fee mechanism design aiming at mitigating blockchain insufficient storage fee issue. The goal of the mechanism is to ensure the long-term stability and security of blockchain system, as the ever-increasing storage costs become a significant burden to miners and reduce their incentives to participate in the blockchain operation.
		\item \textit{Three-Stage Interaction Model:} 
		We propose a three-stage game-theoretical model to characterize the interactions among the protocol designer, users, and miners. 
		The analysis of the model is analytically challenging as the user-miner fee interaction is a two-stage queueing game. Specifically, the state transition of the queue system is stochastically affected by the decision of all users and miners, and each miner faces an integer programming problem which is game-theoretically coupled with other miners' strategies. Nevertheless, we can derive the subgame perfect equilibrium of the model in closed form.
		\item \textit{Explaining the deficiency of the existing protocol:} Through the analysis of three-party interaction, we find that under the existing protocol each miner is unaware of the negative externality that it imposes on other miners when making transaction selections. This causes the miners to accept transactions with fees not enough to cover the overall system storage costs. 
		
		\item \textit{Proposing a mechanism to generate sufficient fee and achieve unconstrained social optimum:} 
		We propose an FWT mechanism motivated by two types of negative externalities in the system. We show that the optimal FWT mechanism incentivizes users to pay sufficient transaction fees for the overall system storage costs while achieving the unconstrained social optimum. Surprisingly, we also find that users who impose lower waiting time costs on other users may pay a higher waiting tax under the optimal FWT mechanism, as the optimal FWT mechanism encourages other users to generate more transactions to maximize the social welfare.
		\item \textit{Ethereum-based numerical results:} 
		We conduct the numerical analysis based on practical Ethereum blockchain environment. Compared with the existing protocol, our proposed optimal FWT mechanism not only produces sufficient transaction fees but also achieves an average social welfare improvement of 33.73\% or more. Moreover, the optimal FWT mechanism achieves the maximum fairness index. Even though we relax the assumption of homogeneous-storage-cost miners, the optimal FWT mechanism still achieves an average social welfare improvement of 61.49\% over the existing protocol.
		\vspace{-3pt}
	\end{itemize}
	
	The rest of the paper is organized as follows. 
	Section \ref{Related_work} reviews the related literature. Section \ref{System_model} introduces the system model. We characterize the mathematical details and derive the closed-form subgame perfect equilibrium of the model's three stages in Sections \ref{Stage_III},\ref{Stage_II}, and \ref{Stage_I}, respectively. We evaluate the system performance in Section \ref{Evaluation}.
	We conclude this paper in Section \ref{Conclusion}. 
	\section{Literature Review}\label{Related_work}
	Our work studies the mechanism design on the user-miner fee interaction regarding blockchain storage. Hence, we review the previous literature from two aspects, i.e., analysis of the fee interaction scheme in the existing blockchain and new fee mechanism design in blockchain. 
	\subsection{Analysis of Fee Interaction Scheme in Existing Blockchain}
	The first group of literature (e.g., \cite{Monopoly_blockchain,Mining_to_market,Transaction_Queuing,Block_size,Feewoblocksize2}) analyzing how users set transaction fees regarding the transaction waiting time in the existing blockchain. Huberman \emph{et al.} in \cite{Monopoly_blockchain} found that the desire to shorten the transaction waiting time is the primary reason for users to pay transaction fees. Following this work, several other papers analyzed how waiting time affects users' transaction fees. Easley \emph{et al.} in \cite{Mining_to_market} showed that as the average transaction waiting time increases, the ratio of users who pay fees also increases. Li \emph{et al.} in \cite{Transaction_Queuing} analyzed the case where users choose between paying a fixed-level fee or not paying the fee. They revealed that an excessive waiting time would discourage low-waiting-time-cost users from paying transaction fees. Several further studies in \cite{Block_size,Feewoblocksize2} investigated the factors in blockchain that may impact the waiting time. These works proved that both the block production time and the block propagation time affect the waiting time (hence the transaction fee). However, the previous works did not consider how miners tradeoff between the transaction fees and storage costs. We consider a general model where each miner chooses the transactions considering the tradeoff between the transaction fees and storage costs. This consideration significantly complicates the analysis.
	
	\subsection{New Fee Mechanism Design in Blockchain}
	The second group of literature focused on transaction fee market design (e.g., \cite{EIP_1559,Fee_design_ABC,Fee_design_Auction,Fee_design_fee_market,Fee_design_re_fee_market}) with different goals. Vitalik \emph{et al.} in \cite{EIP_1559} proposed a burning base fee mechanism to make the fee prediction easy for the users. Some works aimed to improve the system performance. Hu \emph{et al.} in \cite{Fee_design_ABC} proposed a correlated-equilibrium-based fee mechanism to achieve both the individual and global optimum. Ai \emph{et al.} in \cite{Fee_design_Auction} applied the double auction to improve the fairness of the system.
	The literature \cite{Fee_design_fee_market} used the second prize auction to reduce the variance of transaction fees. Lavi \emph{et al.} in \cite{Fee_design_re_fee_market} showed that the monopolistic auction is resilient to market manipulation. Overall, the current work on the fee mechanism design did not consider two types of negative externalities in miners' transaction selections and users' transaction generations. Our work is one of the first analytical studies to explicitly considers the mitigation of two types of negative externalities in the blockchain system.



	\section{System Model}\label{System_model}
	In this section, we describe the system model of the blockchain. We first introduce the high-level operation process of the blockchain system in Section \ref{Blockchain} and then discuss our proposed FWT mechanism in Section \ref{Key_idea}. Finally we propose a three-stage Stackelberg game to characterize the blockchain system in Section \ref{Game_model}. 
	
	\subsection{Blockchain Operation}\label{Blockchain}
	In this subsection, we briefly introduce the operation process of blockchain.
	
	\begin{figure}[!h]
		\vspace{-2mm}
		\centerline{\includegraphics[width=5.2cm]{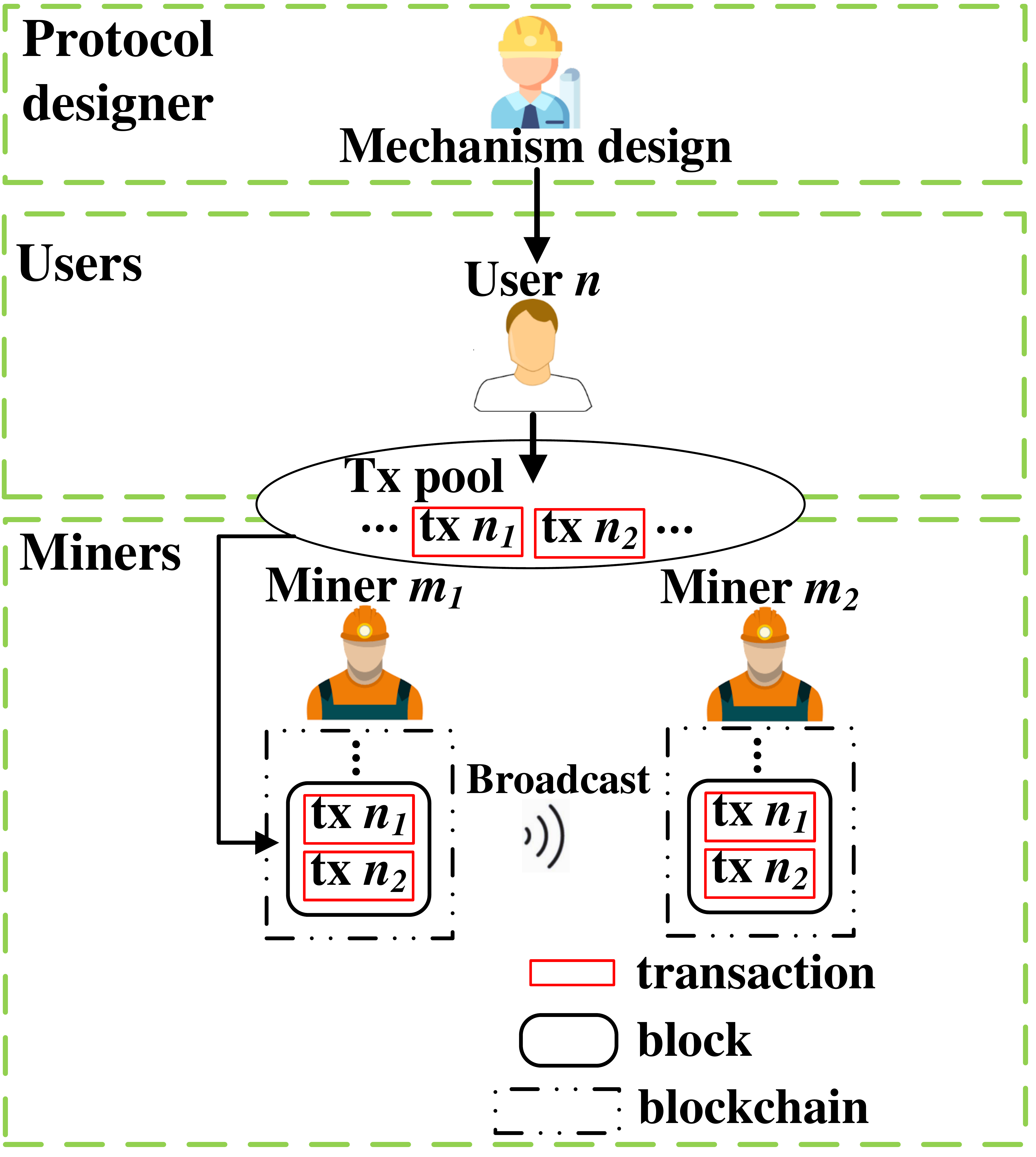}}
		\vspace{0mm}
		\caption{Blockchain operation.}
		\vspace{-3mm}
		\label{BC_process}
	\end{figure}
	
	Fig. \ref{BC_process} illustrates the typical blockchain operation \cite{Bitcoinbook}. The protocol designer first determines the mechanism for users and miners. Then the users generate transactions and choose the transaction fees. Finally, miners select transactions and include them in the blockchain through mining. The details are as follows:
	\begin{enumerate}
		\item Protocol designer's mechanism design: The protocol designer determines the consensus protocol for the system. The blockchain online community usually collectively acts as the protocol designer. For example, in 2018, the Ethereum online community\footnote{\url{https://www.reddit.com/r/ethereum/}} proposed Ethereum Improvement Proposal 1234 to decrease block reward by 33\%.
		\item Users' transaction generation: A user $n$ generates two transactions (tx $n_1$ and tx $n_2$) and assigns the fee-per-byte value for each transaction.\footnote{Users often set the fee-per-byte rather than the transaction fee in Bitcoin \cite{Fee_per_byte}.} The transaction fee of a transaction satisfies: 
		\vspace{-1mm}
		\begin{equation*}
		\text{transaction fee =  transaction size $\times$ fee-per-byte.}
		\vspace{-1mm}
		\end{equation*}
		The transaction fee serves as an incentive for miners to include the transaction into a future \textit{block}.\footnote{A block is a container of transactions. In Bitcoin \cite{Bitcoin_white_paper}, a block contains the cryptographic hash of the previous block, a time-stamp, and the data \cite{cai2018decentralized}.} Each generated transaction enters the transaction pool (tx pool) and waits for miners to include it in the blockchain.
		\item Miners' mining: The process of mining a new block (also referred to as one round of mining) contains several steps, as follows: 
		\begin{itemize}
			\item First, each miner selects a set of transactions from the transaction pool (e.g., miner $m_1$ selects both tx $n_1$ and tx $n_2$). 
			\item Next, miners compete to solve a cryptographic puzzle. Once a miner solves the puzzle (being first among all miners), he will pack his selected transactions, the puzzle solution, along with some auxiliary data into a block. The miner who produces such a new block can get the fees from his selected transactions (e.g., miner $m_1$ gets fees of tx $n_1$ and tx $n_2$) and the block reward (for generating this new block) as a bonus. The transactions in the new block are included in the blockchain.
			\item Finally, the miner who produces the new block broadcasts the block information to his neighbors in the network, and \emph{all} miners need to update the local storage to include the new block. 
		\end{itemize}
	\end{enumerate}
	
	Next, we will introduce our proposed Fee and Waiting Tax (FWT) mechanism for the protocol designer.
	
	\subsection{FWT Mechanism}\label{Key_idea}
	There are two types of negative externalities in the existing blockchain protocol, i.e., each miner's transaction selection imposes storage costs on other miners and each user's transaction generation increases the average waiting time of other users. These two types of negative externalities are why miners accept insufficient-fee transactions and users experience excessive waiting time, respectively, in the existing protocol. Motivated by the negative externalities, we propose the FWT mechanism as follows:
	\begin{enumerate}
		\item Fee choices: The FWT mechanism offers several fee-per-byte choices for users. Users can choose the transaction generation rates for different fee-per-byte choices. The protocol designer properly optimizes the fee-per-byte choices such that the users pay sufficient fees to cover the total system storage costs. 
		\item Waiting tax: The FWT mechanism imposes a waiting tax on each user based on the negative impact that he generates on others, so that the user will be more conservative in generating transactions. 
	\end{enumerate}
	Next, we will present the FWT mechanism in more detail. 
	
	\subsection{Three-stage Stackelberg Game}\label{Game_model}
\begin{figure}[!htbp]
		\centerline{\includegraphics[width=6.1cm]{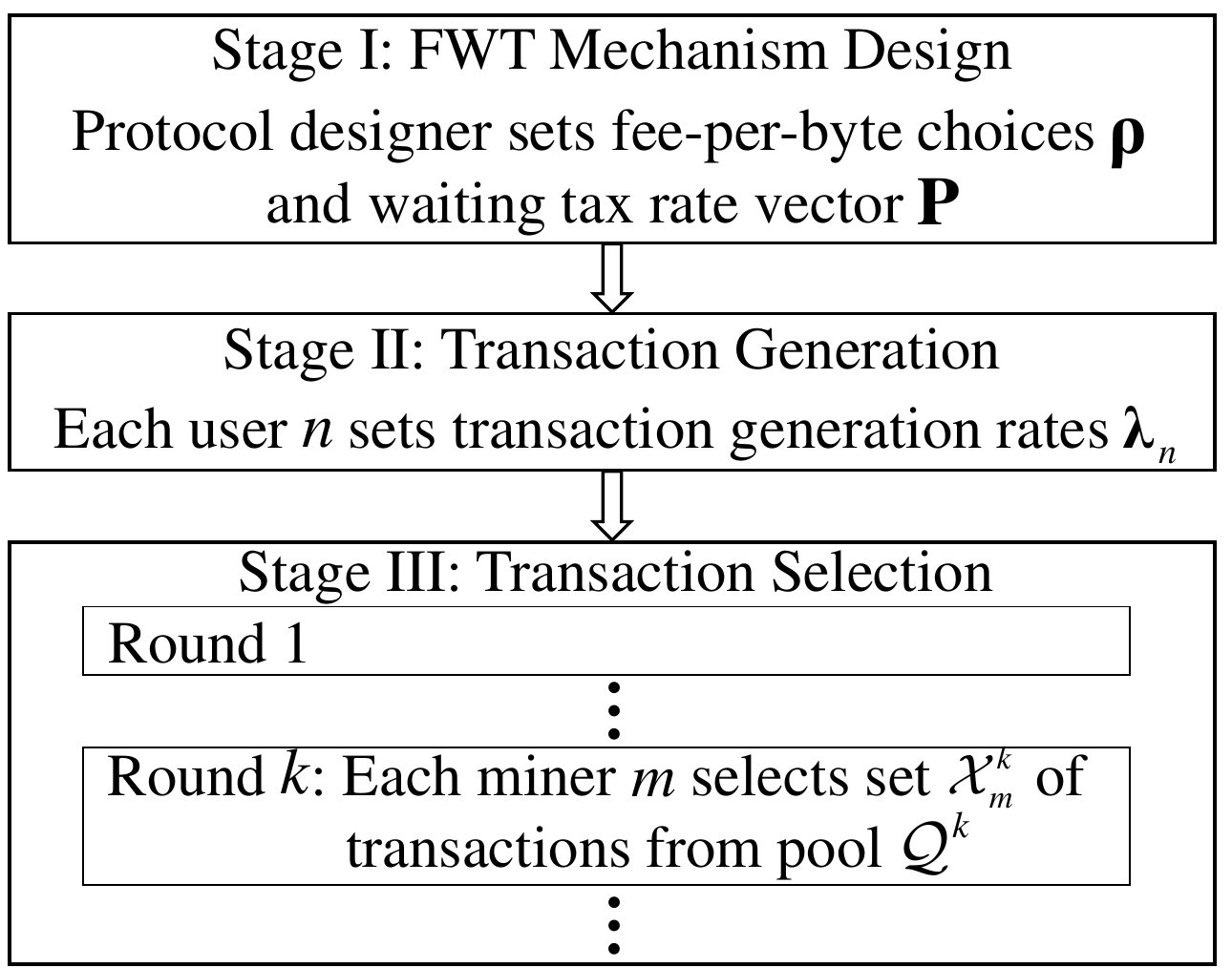}}
		\vspace{-1mm}
		\caption{Three-stage Stackelberg game.}
		\label{Three-stage}
		\vspace{-3.5mm}
	\end{figure}

	We model the interactions among the protocol designer, $N$ users, and $M$ miners as a three-stage Stackelberg game, as illustrated in Fig. \ref{Three-stage}. 
	
	In Stage I, the protocol designer ensures users to pay sufficient fees by setting \textit{fee-per-byte choices} $\boldsymbol{\rho} = (\overline{\rho},\underline{\rho})$. Here we consider the protocol designer offers users two fee-per-byte choices. The binary choice is a simplification, but it can show users' tradeoff between transaction fees and waiting time. Moreover, the protocol designer maximizes the social welfare (a common objective in the literature \cite{Blockchain_SW1}\cite{Blockchain_SW2}) by setting \textit{the waiting tax rate vector} $\boldsymbol{P}$. The waiting tax rate vector specifies each user's tax payment to all other users, compensating the waiting costs that the user imposes on others.
	
	In Stage II, each user $n$ tradeoffs between paying high transaction fees and bearing long time waiting for transaction inclusion. More specifically, the user chooses the transaction generation rates $\boldsymbol{\lambda}_n = (\lambda_{n1},\lambda_{n2})$, which denote the transaction generation rates corresponding to the high and low fee-per-byte choices (i.e., $\overline{\rho}$ and $\underline{\rho}$), respectively. Such a differentiated generation rate and fee-per-byte choice provide flexibility to meet the requirements of different applications.\footnote{For example, in Ethereum, all top-3 users who pay most transaction fees generate transactions with significantly different gas prices (10 times) for different applications simultaneously \cite{Eth_Top_Fee}. } Moreover, the waiting tax rate vector $\boldsymbol{P}$ assigns different taxes to different types of users (the details are in Section \ref{Stage_II_model}) and each user pays the waiting tax to all the others accordingly.
	
	In Stage III, mining proceeds continuously over time. Without loss of generality, we examine the round $k = 1,2,\cdots$ of mining, during which miners mine the block $k$. The length of each round $k$ (the time between the successful mining of block $k-1$ and $k$) follows an exponential distribution.\footnote{The exponential distribution is confirmed by Bitcoin data analysis \cite{BlockGeneration2018} and is also commonly done in blockchain analysis \cite{Mining_to_market}\cite{Gap_Game}.} We further assume that the block propagation delay is zero,\footnote{This is a valid assumption because the average block propagation delay in Bitcoin is roughly 2\% of block interval time \cite{Block_propagation_time}.} i.e., all miners receive the new block as soon as some miner successfully mines such a block.
	When determining what to include in a block, each miner $m$ wants to achieve the proper balance between receiving transaction fees and bearing storage costs. The timeline of round $k$ is as follows:
	\begin{enumerate}[leftmargin=13pt]
		\item First, each miner $m$ selects a set of transactions $\mathcal{X}_{m}^k$ from the transaction pool to include in the new block. The transaction pool is the set of all transactions waiting to be included in a block.
		\item During round $k$, users may generate transactions at any time. The newly generated transactions enter the transaction pool and each miner $m$ can change his transaction selection $\mathcal{X}_{m}^k$.\footnote{The reason is that the mining process is memoryless and the success rate of finding a block is independent of the included transaction \cite{BlockGeneration2018}.} We denote the transaction pool just before miners find block $k$ as $\mathcal{Q}^k$ such that $\mathcal{X}_{m}^k\subseteq\mathcal{Q}^k$ and notice that finding a new block is a stochastic event. 
		\item When a miner finds block $k$, the round $k$ ends. The miner who finds the block $k$ receives the transaction fees from the transactions included in his proposed block. All the miners store the block $k$ and bear the costs of storage individually. The transaction pool updates by removing those transactions that have been included in the block $k$.
	\end{enumerate}

\begin{figure}[!h]
		\vspace{-2mm}
		\centerline{\includegraphics[width=7.6cm]{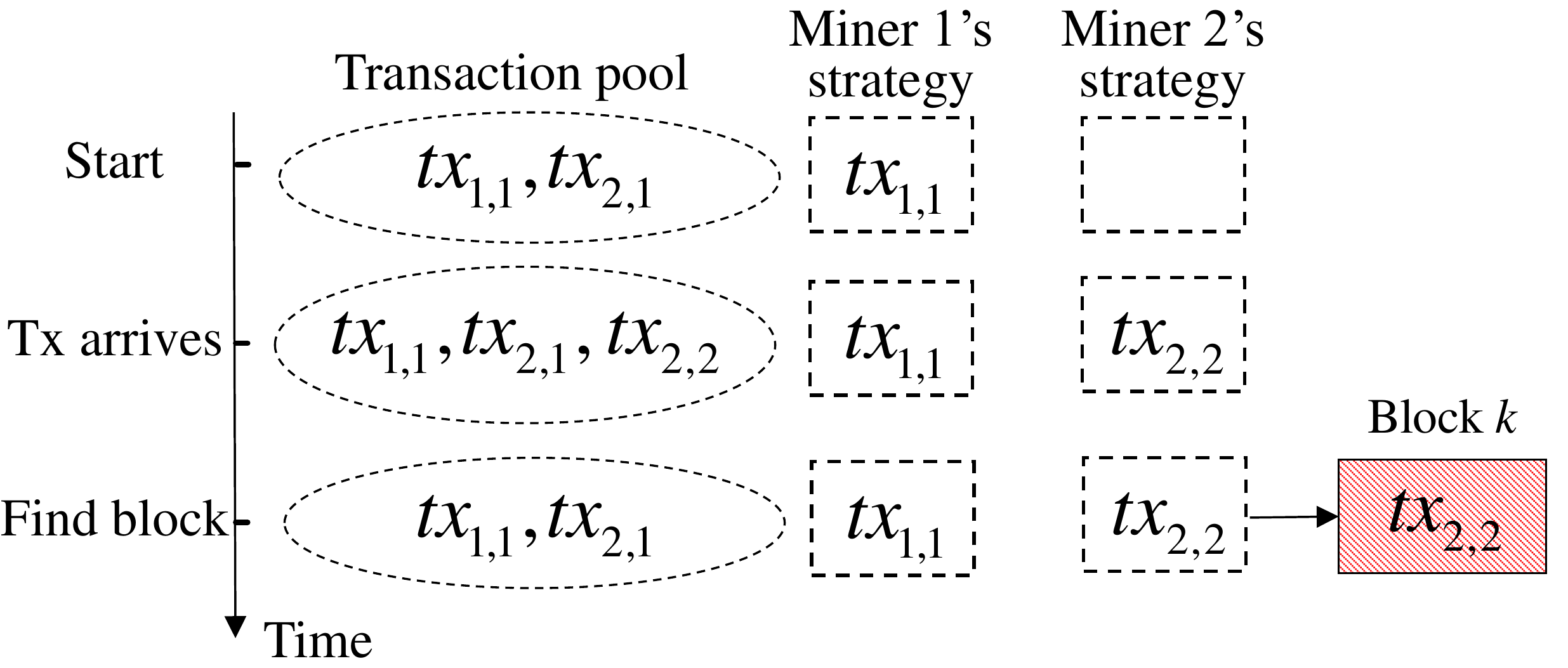}}
		\caption{Timeline of round $k$.}
		\vspace{-2mm}
		\label{Model_timeline}
	\end{figure}

	Fig. \ref{Model_timeline} illustrates the above mining process with the case of 2 users and 2 miners. For multiple transactions generated by user $n$, we will differentiate them in the subscript $i$, i.e., $tx_{n,i}$ $(i=1,2,\cdots)$.
	
	\begin{enumerate}[leftmargin=13pt]
		\item First, there are two transactions $tx_{1,1}$ and $tx_{2,1}$ in transaction pool in Fig. \ref{Model_timeline}. Miner 1 and 2 adopt strategies $\mathcal{X}_1^{k} = \{tx_{1,1}\}$ and $\mathcal{X}_2^{k} = \emptyset$, respectively.
		\item During round $k$, user 2 generates a new transaction $tx_{2,2}$ and it enters transaction pool. Miner 1's strategy remains the same while miner 2 changes his strategy to $\mathcal{X}_2^{k} = \{tx_{2,2}\}$. In this example, $\mathcal{Q}^{k} = \{tx_{1,1},tx_{2,1}, tx_{2,2}\}$.
		\item Miner 2 finds a block $k$ and round $k$ ends. Miner 2 includes $tx_{2,2}$ in blockchain and transaction pool deletes $tx_{2,2}$.
	\end{enumerate}
	
	In the next three sections, we will introduce the mathematical detail of each stage of the model and analyze it through backward induction.

	\section{Stage III: Transaction Selection Equilibrium of Miners}\label{Stage_III}
	In this section, we will characterize how miners select transactions in Stage III. We first model miners' transaction selections in round $k =1,2,\cdots$ of mining as a game in Section \ref{Stage_III_model}, then we characterize the Nash equilibrium of the game in Section \ref{Stage_III_Analysis}.

	\subsection{Model of Miners Transaction Selection in Round $k$}\label{Stage_III_model}Although miners will interact with each other over many rounds of mining, we will focus on a particular round $k$ of mining, during which each miner selects a set of transactions to maximize his own payoffs.\footnote{Such a myopic setting is widely used in blockchain analysis \cite{Gap_Game}\cite{Myopic_mining}. It is a typical setting in dynamic games with many players, where each player's influence is small.} We formulate the miners' interaction as a non-cooperative game. 
	
	\subsubsection{Miners}We consider the set of miners as $\mathcal{M} = \{1,\cdots,M\}$.\footnote{We analyze the system in a quasi-static state \cite{Mining_to_market}\cite{Gap_Game}\cite{Fruitchains}. That is to say, there are no users or miners joining or leaving the system.} The normalized mining power (e.g., computing power in proof of work) of miner $m\in\mathcal{M}$ is $\alpha_m>0$, which represents the probability of miner $m$ successfully finding a block. We have $\sum_{m\in\mathcal{M}}\alpha_m = 1$.
	
	\subsubsection{Miners' strategies}
	Each miner $m$ selects a set $\mathcal{X}_m^{k} \subseteq\mathcal{Q}^k$ of transactions from the transaction pool $\mathcal{Q}^k$. For the ease of analysis, we assume that a block can contain at most one transaction.\footnote{This assumption characterizes the reality of block size limit, which has been adopted in \cite{Mining_to_market}\cite{Transaction_Queuing}. Blockchain platform IOTA adopts the one transaction per block rule \cite{IOTA}. Considering multiple transactions per block means that each miner's strategy space becomes multi-dimensional, making the problem much more challenging to solve. We will relax this assumption in future work.} Thus miner $m$'s strategy satisfies $|\mathcal{X}_m^{k}|\in\{0,1\}$. As transactions vary in both sizes and fees, we adopt a benign assumption where each miner either selects the highest fee-per-byte transaction from the transaction pool or selects no transaction, which is consistent with the empirical studies \cite{Fee_per_byte}\cite{Fee_per_byte1}. Existing fee recommendation softwares also follow this rule \cite{Fee_recommendation1}\cite{Fee_recommendation2}.
	
	To simplify the description, we use $\mathcal{X}_m^k = \{(n,i)\}$ (or $\emptyset$) to represent that miner $m$ selects $tx_{n,i}$ (or no transaction) in round $k$.
	
	\emph{Miners' payoff functions: }Miner $m$'s payoff depends on both the transaction fee and storage cost.\footnote{We do not consider the block reward and the cost of running the mining machine since they are not affected by each miner's transaction selection. Besides, the transaction fees are the key to cover blockchain storage costs as the block reward gradually shrinks.}
	
	\begin{itemize}
		\item Transaction fee: For a transaction $tx_{n,i}$, its transaction fee is the product of the transaction size and fee-per-byte, i.e., $s_{n,i}\rho_{n,i}$. Only the miner who successfully finds a block receives the transaction fees from his selection. Thus miner $m$ will get the total transaction fees of $\sum_{(n,i) \in {\mathcal{X}_m^k}} {s_{n,i}\rho_{n,i}}$ with probability $\alpha_m$.
		\item Storage cost: For analysis, we assume that all miners have homogeneous storage cost of $C_s$ per byte, representing that miners use similar storage technology. We will consider the impact of the heterogeneous storage costs numerically in Section \ref{Num2}. Storing a transaction $tx_{n,i}$ with size $s_{n,i}$ imposes a storage cost $s_{n,i}C_s$ to a miner. If any miner $j\in \mathcal{M}$ selects transaction $tx_{n,i}$ (i.e., $\mathcal{X}_{j}^{k} = \{(n,i)\}$) and successfully finds a block (with probability $\alpha_j$), all miners need to store that block and bear the storage costs $s_{n,i}C_s$ for the transaction $tx_{n,i}$ that miner $j$ selects \cite{Blockchain_survey}. Overall, miner $m$ will bear the storage costs of 
		\begin{equation}\label{Storage_cost}
		C^k(\mathcal{X}_m^{k},\boldsymbol{\mathcal{X}}_{-m}^{k}) = \sum\limits_{j\in\mathcal{M}}\alpha_j\sum\limits_{(n,i) \in {\mathcal{X}_j^k}} s_{n,i}C_s
		\end{equation}
		in round $k$,\footnote{We neglect the storage costs of non-transaction data since it is very small. For example, on June 2nd, 2020, the average block size of Bitcoin is 1.284 MB and non-transaction data in a block takes up less than 1 KB. https://www.blockchain.com/charts/avg-block-size.} where $\boldsymbol{\mathcal{X}}_{-m}^{k} = (\mathcal{X}_j^{k},\forall j\in\mathcal{M},j\not=m)$ represents the strategies of all the miners other than $m$. Miner $m$'s storage costs in (\ref{Storage_cost}) reveals the \emph{negative externality in transaction selection}: when a miner selects a transaction and finds a block, it imposes storage costs to all the other miners.
	\end{itemize}
	Combining the transaction fee and storage cost, miner $m$'s payoff in round $k$ is: 
	\begin{equation}\label{Miner_utility}
	v_{m}^{k}(\mathcal{X}_m^{k},\boldsymbol{\mathcal{X}}_{-m}^{k},\boldsymbol{\rho}) = \alpha_m\sum\limits_{(n,i) \in {\mathcal{X}_m^k}} {s_{n,i}\rho_{n,i}}-C^k(\mathcal{X}_m^{k},\boldsymbol{\mathcal{X}}_{-m}^{k}).
	\end{equation}
	
	
	\subsubsection{Game formulation} We formulate the round $k$ of mining as a non-cooperative game, where miners simultaneously select the transactions (to be included in his block) to maximize their own payoffs.
	\begin{gam}[Stage III: Transaction Selection Game in round $k$]\label{Stage_III_game} In Stage III, Transaction Selection Game in round $k =1,2,\cdots$ is a tuple $\Phi^k =(\mathcal{M},\mathcal{X}^k, \boldsymbol{V}^k)$ defined by:
		\begin{itemize}
			\item Players: The set of miners $\mathcal{M}$. 
			\item Strategies: Each miner $m\in \mathcal{M}$ selects a set $\mathcal{X}_m^k\in\mathcal{B}_m^k=\{\mathcal{A}|\mathcal{A}\subseteq\mathcal{Q}^k, |\mathcal{A}|\in\{0,1\}\}$ of transactions. The strategy profiles of all the miners is $(\mathcal{X}_m^k,\forall m\in\mathcal{M})$. The set of feasible strategy profile of all miners is $\mathcal{X}^k = \prod_{m \in \mathcal{M}} \mathcal{B}_m^k$.
			\item Payoffs: The vector $\boldsymbol{V}^k = (v_{m}^k,\forall m\in M)$ contains all miners' payoffs as defined in (\ref{Miner_utility}).
		\end{itemize}
	\end{gam}
	In Game \ref{Stage_III_game}, each miner tradeoffs between the transaction fee and storage cost to maximize his payoff, considering the strategies of other miners. Specifically, on the one hand, miner $m$ gets high revenue for selecting a high-fee transaction and finding a block. Meanwhile, for the highest-fee-per-byte transaction, if its fee is lower than its storage cost, a miner may still select it if all the other miners select it and he will eventually bear the storage cost of it.

	\subsection{Nash Equilibrium Analysis}\label{Stage_III_Analysis}
	We first define the Nash equilibrium in Definition \ref{NE_define}.
	\begin{defin}[Nash Equilibrium]\label{NE_define}
		Given the fee-per-byte choices $\boldsymbol{\rho}$, a strategy profile $(\mathcal{X}_m^{k, {\rm NE}},$ $\forall m \in \mathcal{M})$ constitutes a \emph{Nash equilibrium} in Game 1 if
		\begin{equation}
		\begin{aligned}
		&v_m^{k}(\mathcal{X}_m^{k,{\rm NE}},\boldsymbol{\mathcal{X}}_{-m}^{k,{\rm NE}},\boldsymbol{\rho}) \geqslant v_m^{k}(\mathcal{X}_m^{k},\boldsymbol{\mathcal{X}}_{-m}^{k,{\rm NE}},\boldsymbol{\rho}), \\
		&\forall \mathcal{X}_m^{k}\in\mathcal{B}_m^k, \forall m \in \mathcal{M}.
		\end{aligned}
		\end{equation}
	\end{defin}
	
	We define some strategy-related notation for the ease of presentation. When the transaction pool in round $k$ is not empty, the set of all the highest-fee-per-byte transactions in the transaction pool as $\mathcal{Q}^{k,\text{high}} \triangleq \{(n,i)\in\mathcal{Q}^{k}|\rho_{n,i}\geqslant\rho_{j,l},\forall (j,l)\in\mathcal{Q}^k\}.$ Within the set $\mathcal{Q}^{k,\text{high}}$, we define the transaction with the earliest generation time is
	\begin{equation}
	(n^{k*},i^{k*}) \triangleq \underset{(n,i)\in\mathcal{Q}^{k,\text{high}}}{\arg \min } \hspace{1mm}t^{\rm gen}_{n,i},
	\end{equation}
	where $t^{\rm gen}_{n,i}$ is the generation time of transaction $(n,i)$.
	Then, we summarize the Nash equilibrium as follows.
	
	\begin{thm}[Miners' Equilibrium in Stage III]\label{Stage_III_NE}
		The strategy profile $(\mathcal{X}_m^{k, {\rm NE}}, \forall m \in \mathcal{M})$ constitutes a Nash equilibrium in round $k$, where 
		$$\mathcal{X}_m^{k, {\rm NE}} = \begin{cases}
		\{(n^{k*},i^{k*})\},&\hspace{-2mm}\text{if }\mathcal{Q}^{k}\not=\emptyset\text{ and }\rho_{n^{k*},i^{k*}} \geqslant C_s,\\
		\emptyset,&\hspace{-2mm}\text{otherwise.}
		\end{cases}$$
	\end{thm}
	Due to the space limit, we leave the proofs of all mathematical results in the online appendix \cite{Online_Appendix}.
	
	Corollary 1 reveals an interesting observation from Theorem \ref{Stage_III_NE}.
	\begin{cor}
		Each miner only accepts $tx_{n,i}$ if its fee-per-byte is higher than a miner's storage cost per byte, i.e., $\rho_{n,i}\geqslant C_s$. However, a miner's storage cost per byte $C_s$ is insufficient to cover all miners' total storage cost per byte, i.e., $MC_s$.
	\end{cor}
	Corollary 1 mathematically reveals the \emph{negative externality in transaction selection} introduced in Section \ref{Stage_III_model}. Each miner only considers his own storage cost when selecting the transaction, without considering the negative impact on all other miners in the system. Hence even if the transaction fee can cover the storage cost of a single miner, it can be far from enough to cover the total storage cost of system. As miners accept insufficient-fee transactions, users may not pay enough transaction fees to cover all miners' total storage costs, causing the storage sustainability issue.
	
	\section{Stage II: Transaction Generation Equilibrium of Users}\label{Stage_II}
	In this section, we will characterize how users generate transactions in Stage II. We first formulate users' transaction generation as a game in Section \ref{Stage_II_model}, then we characterize the Nash equilibrium of the game in Section \ref{Stage_II_Analysis}.
	\subsection{Model of Users Transaction Generation}\label{Stage_II_model}In Stage II, users set the \emph{transaction generation rates} to maximize their own payoffs. 
	\subsubsection{Users' strategies} We denote the set of users as $\mathcal{N} = \{1,\cdots,N\}$. Given two fee-per-byte choices $\overline{\rho}$ and $\underline{\rho}$ offered by the protocol designer, each user $n\in\mathcal{N}$ generates transactions at each choice following a Poisson process. The strategy of user $n$ is to set the \emph{transaction generation rates} ${\boldsymbol{\lambda}_n} = (\lambda_{n1},\lambda_{n2})$, where $\lambda_{n1}$ and $\lambda_{n2}$ are the rates of user $n$ generating transactions at $\overline{\rho}$ and $\underline{\rho}$, respectively, satisfying the following constraints: 
	\begin{equation}\label{Tx_generation_constraint}
	\begin{cases}
	\lambda_{n1}+\lambda_{n2} \leqslant \frac{\mu}{N},\\
	\lambda_{n1},\lambda_{n2}\geqslant 0,
	\end{cases}
	\end{equation}
	where $\mu$ is the system average block generation rate and each user's maximum transaction generation rate is $\frac{\mu}{N}$. Constraint (\ref{Tx_generation_constraint}) ensures that it is feasible to include all generated transactions from all users in the blockchain (if the miners choose to do so in Stage III).
	
	\subsubsection{Transaction waiting time} 
	Here we define the waiting time of any transaction $tx_{n,i}$ as the time lapse between the generation time and the on-chain time.
	\begin{itemize}
		\item Generation time of transaction $tx_{n,i}$ is denoted as $t^{\rm gen}_{n,i}$.
		\item On-chain time: When a miner selects transaction $tx_{n,i}$ and finds a block in round $k(=1,2,\cdots)$, then round $k$ ends, and transaction $tx_{n,i}$ is included in the blockchain. Thus, round $k$'s ending time $t^{\rm end}(k)$ is the transaction on-chain time, i.e.,
		\begin{equation}
		\hspace{-4mm}t^{\rm on}_{n,i} = \begin{cases}
		t^{\rm end}(k), &  \hspace{-2mm}\text{if $tx_{n,i}$ is included in block $k$,}\\
		\infty, & \hspace{-2mm}\text{if $tx_{n,i}$ is not included in any block.}
		\end{cases}
		\end{equation}
		\item Waiting time is the difference between the on-chain time minus and the generation time, i.e., $w_{n,i} = t^{\rm on}_{n,i}-t^{\rm gen}_{n,i}$. Waiting time $w_{n,i}$ is a random variable as the block generation is stochastic. The rate of transactions entering the transaction pool affects the waiting time $w_{n,i}$, thus it is a function of all users transaction generation rates, i.e., $\boldsymbol{\lambda} = (\boldsymbol{\lambda}_n,\forall n\in\mathcal{N})$. We will compute the expectation of $w_{n,i}$ in Lemma \ref{Waiting_time} of Section \ref{Stage_II_Analysis}.
	\end{itemize}
	
	\emph{Negative externality in transaction generation}: When user $n$ generates a transaction and miners include it in the blockchain, other transactions in the transaction pool have to wait. Thus, user $n$'s transactions increase the average waiting time of all the other users' transactions. If a user maximizes his own payoff without considering the negative externality, all the other users will experience excessive waiting time. This motivates us to propose the waiting tax to let each user internalize such a negative externality.
	
	\subsubsection{User $n$'s surplus obtained from one transaction $tx_{n,i}$} User $n$'s surplus obtained from one transaction $tx_{n,i}$ depends on whether $tx_{n,i}$ is included in the blockchain.
	\begin{itemize}[leftmargin=12pt]
		\item If $tx_{n,i}$ is included in the blockchain: The surplus depends on the on-chain utility from one transaction, transaction fee, waiting time cost, and waiting tax. 
		\begin{itemize}[leftmargin=10pt]
			\item User $n$'s on-chain utility from $tx_{n,i}$: When $tx_{n,i}$ is included in the blockchain, user $n$ will obtain utility of $R_n$. For example, a user gets a certain level of utility when successfully purchasing a kitty in Ethereum-based game cryptokitties. To model the users' heterogeneity of utilities, we consider two user types: with $N_H$ high-utility users (type-$H$) and $N_L = N-N_H$ low-utility users (type-$L$).\footnote{Two applications can capture heterogeneous users set transaction fee to compete for shorter waiting time.} Thus, user $n$'s on-chain utility from one transaction is
			\vspace{-1mm}
			\begin{equation}\label{Tx_recorded_utility}
			\hspace{-3mm}R_n = \begin{cases}
			R_H, & \hspace{-2.5mm}\text{if user $n$ is type-$H$,}\\
			R_L, & \hspace{-2.5mm}\text{if user $n$ is type-$L$,}\\
			\end{cases}
			\vspace{-1mm}
			\end{equation}
			where $R_H\geqslant R_L$. Our model generalizes the homogeneous utility model in \cite{Mining_to_market} and \cite{Transaction_Queuing}.
			\item Transaction fee of $tx_{n,i}$: User $n$ pays the transaction fee $f_{n,i}= s_{n,i}\rho_{n,i}$ to the miner who includes $tx_{n,i}$ in the blockchain. We model the size of any transaction follows an i.i.d. distribution with the expectation of $\bar{s}$. The fee-per-byte $\rho_{n,i}$ belongs to the protocol designer's fee-per-byte choices, i.e., $\rho_{n,i} \in\{\overline{\rho},\underline{\rho}\}$. 
			\item Waiting time cost of $tx_{n,i}$: The transaction waiting time $w_{n,i}(\boldsymbol{\lambda})$ imposes a cost to user $n$, which we assume to be a linear function with the impatience coefficient $\gamma$, i.e., $\gamma w_{n,i}(\boldsymbol{\lambda})$. A higher $\gamma$ means users are less patient.
			\item Waiting tax of $tx_{n,i}$: Since user $n$'s transaction generation increases the expected transaction waiting time of any other user $l\not=n$, we introduce the waiting tax $p_{nl}$ to internalize this negative externality. More specifically, user $n$ pays user $l$ the amount of $p_{nl}$ to compensate the waiting costs $n$ imposes. Depending on the types of users $n$ and $l$, the possible waiting tax has four different values $\boldsymbol{P} = (P_{HH}, P_{HL}, P_{LH},P_{LL})$:
			\begin{equation}\label{Waiting_time_price}
			p_{nl} = \begin{cases}
			P_{HH}, &\hspace{-3mm}\text{if both user $n$ and $l$ are type-$H$,}\\
			P_{HL}, &\hspace{-3mm}\text{if user $n$ is type-$H$ and user $l$ is type-$L$,}\\
			P_{LH}, &\hspace{-3mm}\text{if user $n$ is type-$L$ and user $l$ is type-$H$,}\\
			P_{LL}, &\hspace{-3mm}\text{if both user $n$ and $l$ are type-$L$.}\\
			\end{cases}
			\vspace{-1mm}
			\end{equation}	
		\end{itemize}
		To sum up, user $n$'s surplus when $tx_{n,i}$ is included in the blockchain is 
		\vspace{-1mm}
		\begin{equation}\label{Tx_recorded_payoff}
		\theta_{n,i}(\boldsymbol{\lambda},\boldsymbol{\rho},\boldsymbol{P})  = R_n-s_{n,i}\rho_{n,i}-\gamma w_{n,i}(\boldsymbol{\lambda})-\sum\limits_{l\in\mathcal{N},l\not=n}p_{nl}.
		\vspace{-1mm}
		\end{equation}
		\item If $tx_{n,i}$ is not included in the blockchain (i.e., not in any block), user $n$ will not get the transaction on-chain utility $R_n$, and he will not pay the fee $f_{n,i}$ or the waiting tax. However, user $n$ still experiences the (possibly infinite) waiting time to know that the transaction will not be included. In this case, user $n$'s surplus from transaction $tx_{n,i}$ is
		\vspace{-1mm}
		\begin{equation}\label{Tx_unrecorded_payoff}
		\theta_{n,i}(\boldsymbol{\lambda},\boldsymbol{\rho},\boldsymbol{P}) = -\gamma  w_{n,i}(\boldsymbol{\lambda}).
		\vspace{-2mm}
		\end{equation}
		
	\end{itemize}
	To simplify the formulation, we define the indicator function to indicate whether $tx_{n,i}$ is included in the blockchain as follows 
	\vspace{-1mm}
	\begin{equation}
	\hspace{-0.6mm}\mathbf{1}(n,i) = \begin{cases}
	\hspace{-0.3mm}1, \text{if $tx_{n,i}$ is included in blockchain,}\\
	\hspace{-0.3mm}0, \text{if $tx_{n,i}$ is not included in blockchain.}
	\end{cases}
	\vspace{-1mm}
	\end{equation}
	Hence user $n$'s surplus obtained from $tx_{n,i}$ can be written as 
	\vspace{-1mm}
	\begin{equation}\label{Theta_n_k}
	\begin{aligned}
	\theta_{n,i}(\boldsymbol{\lambda},\boldsymbol{\rho},\boldsymbol{P})  = &\mathbf{1}(n,i)(R_n-s_{n,i}\rho_{n,i}-\sum\limits_{l\in\mathcal{N},l\not=n}p_{nl})\\
	&-\gamma w_{n,i}(\boldsymbol{\lambda}).
	\end{aligned}
	\vspace{-1mm}
	\end{equation}
	
	\subsubsection{Users' time-average payoff} User $n$'s payoff is the summation of the surplus from all his transactions and the waiting tax paid to him by other users. For user $n$, we denote the number of all his generated transactions in time interval $[0,t]$ as $TX_n(t)$. His time-average payoff is
	\begin{equation}\label{User_utility_fun}
	\begin{aligned}
	&u_n(\boldsymbol{\lambda},\boldsymbol{\rho},\boldsymbol{P})  \\
	=&\lim\limits_{t\rightarrow \infty}\frac{\sum\limits_{i=1}^{TX_n(t)} \mathbb{E}[\theta_{n,i}(\boldsymbol{\lambda},\boldsymbol{\rho},\boldsymbol{P})]+\sum\limits_{l\in\mathcal{N},l\not=n}\sum\limits_{i=1}^{TX_l(t)}\mathbf{1}(l,i)p_{ln}}{t},
	\end{aligned}
	\vspace{0mm}
	\end{equation}
	where $\mathbb{E}[\theta_{n,i}(\boldsymbol{\lambda},\boldsymbol{\rho},\boldsymbol{P})]$ is user $n$'s expected surplus from transaction $tx_{n,i}$. The expectation is taken in terms of random variables transaction size $s_{n,i}$ and waiting time $w_{n,i}$.
	
	\subsubsection{Game formulation} We formulate users' transaction generation as a non-cooperative game, where users set transaction generation rates simultaneously to maximize their own payoffs.\footnote{Here we assume that each user does not consider the influence of his strategic decision on other users (i.e., each user is a price taker). This assumption holds for a blockchain system with many users.}
	\begin{figure*}

		\begin{equation*}
		\lim\limits_{t\rightarrow \infty}\frac{\sum\limits_{i=1}^{TX_n(t)}\mathbb{E}[w_{n,i}(\boldsymbol{\lambda})]}{t} = \begin{cases}
		\frac{\lambda_{n1}}{\mu-\sum_{l\in\mathcal{N}}\lambda_{l1}}+\frac{\mu\lambda_{n2}}{(\mu-\sum_{l\in\mathcal{N}}\lambda_{l1})[\mu-\sum_{l\in\mathcal{N}}(\lambda_{l1}+\lambda_{l2})]}, &\hspace{-2mm}\text{if $C_s\leqslant \underline{\rho}<\overline{\rho}$,\hspace{42.3mm}(14a)} \\
		\frac{\lambda_{n1}}{\mu-\sum_{l\in\mathcal{N}}\lambda_{l1}}, &\hspace{-2mm}\text{if $\underline{\rho}\leqslant C_s<\overline{\rho}$ and $\lambda_{n2} = 0$,\hspace{22.8mm}(14b)}\\
		\infty, &\hspace{-2mm}\text{if $\underline{\rho}\leqslant C_s<\overline{\rho}$ and $\lambda_{n2} > 0$,\hspace{22.8mm}(14c)}\\
		\infty, &\hspace{-2mm}\text{if $\underline{\rho}<\overline{\rho}<C_s$ and ($\lambda_{n1} > 0$ or $\lambda_{n2} > 0$),\hspace{3.3mm}(14d)}\\
		0, &\hspace{-2mm}\text{if $\underline{\rho}<\overline{\rho}<C_s$ and ($\lambda_{n1} = 0$ and $\lambda_{n2} = 0$). (14e)}\\
		\end{cases}
		\end{equation*}
		\vspace{-5mm}
	\end{figure*}
	\begin{gam}[Stage II: Transaction Generation Game]\label{Stage_II_game} In Stage II, Transaction Generation Game is a tuple $\Omega=(\mathcal{N}, \Lambda, \boldsymbol{U})$ defined by:
		\begin{itemize}
			\item Players: The set of users $\mathcal{N}$. 
			\item Strategies: Each user $n$ sets transaction generation rate $\boldsymbol{\lambda}_n$, where the strategy space is $\Lambda_n = \{\boldsymbol{\lambda}_n = (\lambda_{n1},\lambda_{n2})|(\lambda_{n1},\lambda_{n2})\text{ satisfies (\ref{Tx_generation_constraint})}\}$. The strategy profiles of all the users is $\boldsymbol{\lambda} = (\boldsymbol{\lambda}_n,\forall n\in\mathcal{N})$ and the set of all feasible strategy profiles is $\Lambda = \Lambda_1\times\cdots\times\Lambda_N$.
			\item Payoffs: The vector $\boldsymbol{U} = (u_n,\forall n\in \mathcal{N})$ contains all users' payoffs as defined in (\ref{User_utility_fun}).
		\end{itemize}
	\end{gam}
	In Game \ref{Stage_II_game}, each user faces a \emph{tradeoff} between paying a high fee and suffering a high transaction waiting time. Since miners prefer to include transactions with high fees, user $n$ will experience a lower average waiting time by generating more high-fee transactions. However, if paying a high fee is too costly, user $n$ would be better off by generating more low-fee transactions and bearing a higher average waiting time.
	
	\subsection{Nash Equilibrium Analysis}\label{Stage_II_Analysis}
	
	Based on the equilibrium of Stage III, we analyze the equilibrium of Stage II in this subsection. We first compute the transaction waiting time, then we present the users' equilibrium in Stage II.
	\subsubsection{Transaction waiting time}
	According to miners' equilibrium strategies in Stage III, the process of transaction arriving (i.e., users generating transactions) and leaving (i.e., miners including transactions in blockchain) is a two-class M/M/1 queue \cite{queue_book}, where transactions with higher fee-per-byte has priority over transactions with lower fee-per-byte. 
	
	Based on the expected waiting time of two-class M/M/1 queue \cite{queue_book}, we summarize user $n$'s time-average transaction waiting time in following Lemma \ref{Waiting_time}.
	\begin{lem}[Users' Transaction Waiting Time]\label{Waiting_time}
		The time-average transaction waiting time of each user $\forall n\in\mathcal{N}$ is in (14).
	\end{lem}
	For (14a) and (14b), they correspond to the case where miners (eventually) choose to include the transaction as the transaction's fee-per-byte is higher than $C_s$. For (14c) and (14d), they correspond to the case where the transaction waiting time is infinity, as no miner chooses to include the transaction with fee-per-byte strictly lower than $C_s$.
	
	\setcounter{equation}{14}
	
	\subsubsection{Users' Equilibrium in Stage II}
	Here we characterize the users' equilibrium strategies. Similar to prior blockchain literature \cite{Mining_to_market}\cite{liueconomics}, we consider the symmetric Nash equilibrium (SNE) where the same type of users adopt the same strategy. 
	
	For the ease of exposition, we first define some terminology related to the users' equilibrium.\footnote{There can be other SNE but we pay attention to the Pareto-dominant one, where each user achieves no smaller payoff compared to other possible SNEs \cite{Mining_to_market}\cite{huang2021eliciting}.}

	\begin{defin}[Stage II Equilibrium Types]\label{NE_type}\qquad
		\begin{enumerate}
			\item At a $\overline{\rho}$-SNE, all users only generate transactions with the fee-per-byte $\overline{\rho}$.
			\item At a $\underline{\rho}$-SNE, all users only generate transactions with the fee-per-byte $\underline{\rho}$.
		\end{enumerate}
	\end{defin}
	At an equilibrium, each user $n$'s \emph{net transaction utility} $h_n$ plays an important role in his transaction generation rate, which defined as follows
	\begin{equation}\label{Net_utility}
	\hspace{-1mm}h_n = \begin{cases}
	h_H = R_H-[(N_H-1)P_{HH}+N_LP_{HL}], &\hspace{-3.5mm}\text{if $n\in\mathcal{N}_H$,}\\
	h_L = R_L-[N_HP_{LH}+(N_L-1)P_{LL}], &\hspace{-3.5mm}\text{if $n\in\mathcal{N}_L$,}\\
	\end{cases}
	\end{equation}
	where $\mathcal{N}_H$ and $\mathcal{N}_L$ are the set of types $H$ and $L$ users, respectively.
	
	\begin{figure*}
		\begin{equation*}
		\pi_B(h_B,h_S,\rho) = \begin{cases}
		0, &\hspace{17mm} \text{if $h_B\leqslant \bar{s}\rho+\frac{\gamma}{\mu}$, \hspace{0.5mm}(16a)}\\
		A_1(h_B,\rho), &\hspace{-28mm} \text{if $h_B> \bar{s}\rho+\frac{\gamma}{\mu}$ and $h_{S}\leqslant \bar{s}\rho+ \frac{\gamma}{\mu-N_BA_1(h_B,\rho)}$, \hspace{0.5mm}(16b)}\\
		\min\{\frac{\mu}{N_B+N_S},\frac{[\mu-A_2(h_B,h_S,\rho)][(h_B-\bar{s}\rho)A_2(h_B,h_S,\rho)-\gamma]}{N_B[(h_B-\bar{s}\rho)A_2(h_B,h_S,\rho)-\gamma]+N_{S}[(h_{S}-\bar{s}\rho)A_2(h_B,h_S,\rho)-\gamma]}\}, &\hspace{-1mm} \text{if $h_{S}> \bar{s}\rho+ \frac{\gamma}{\mu-N_BA_1(h_B,\rho)}$. \hspace{0.5mm}(16c)}
		\end{cases}
		\end{equation*}
		\begin{equation*}
		\hspace{11mm}\pi_{S}(h_B,h_S,\rho) = \begin{cases}
		0, &\hspace{17.5mm} \text{if $h_B\leqslant \bar{s}\rho+\frac{\gamma}{\mu}$, \hspace{7.5mm}(17a)}\\
		0, &\hspace{-27.5mm} \text{if $h_B> \bar{s}\rho+\frac{\gamma}{\mu}$ and $h_{S}\leqslant \bar{s}\rho+ \frac{\gamma}{\mu-N_BA_1(h_B,\rho)}$,  \hspace{7.5mm}(17b)}\\
		\frac{\mu-N_B\pi_{B}(h_B,h_S,\rho)}{N_{S}}\\
		-\frac{\gamma(N_{S}-1)+\sqrt{\gamma^2(N_{S}-1)^2+4N_{S}(h_{S}-\bar{s}\rho)\gamma[\mu-N_B\pi_{B}(h_B,h_S,\rho)]}}{2(h_{S}-\bar{s}\rho)N_{S}^2}, & \text{if $h_{S}> \bar{s}\rho+ \frac{\gamma}{\mu-N_BA_1(h_B,\rho)}$.  \hspace{7.5mm}(17c)}
		\end{cases}
		\end{equation*}
		\setcounter{equation}{17}
		\begin{equation}\label{A1}
		A_1(h_B,\rho) = \min\{\frac{\mu}{N_B}-\frac{\gamma(N_B-1)+\sqrt{\gamma^2(N_B-1)^2+4\gamma\mu N_B(h_B-\bar{s}\rho)}}{2N^2_B(h_B-\bar{s}\rho)},\frac{\mu}{N_B+N_S}\}.
		\end{equation}
		\begin{equation}\label{A2}
		A_2(h_B,h_S,\rho) = \frac{\gamma(N_B+N_S-1)+\sqrt{\gamma^2(N_B+N_S-1)^2+4\gamma\mu [N_B(h_B-\bar{s}\rho)+N_S(h_S-\bar{s}\rho)]}}{2[N_B(h_B-\bar{s}\rho)+N_S(h_S-\bar{s}\rho)]}.
		\end{equation}
		\vspace{-5mm}
	\end{figure*}
	
	Notice that $h_H$ may not be larger than $h_L$ due to the waiting time tax rate vector $(P_{HH}, P_{HL}, P_{LH},P_{LL})$. We define type-$B$ as the bigger net transaction utility user type (i.e., $B =  {\arg\max}_{l\in\{L,H\}} \hspace{1mm}h_l$) and type-$S$ as the smaller net transaction utility user type (i.e., $S = {\arg \min}_{l\in\{L,H\}}\hspace{1mm} h_l$). We illustrate the connections between types $B$ and $S$ as well as types $H$ and $L$ in Table \ref{table1}. 
	
	\begin{table}[h]\normalsize 
		\centering  
		\caption{Types $B$ and $S$ and corresponding types $H$ and $L$.}
		\label{table1}
		\begin{tabular}{|c|c|c|}  
			\hline
			\text{If $h_H\geqslant h_L$} & $B = H, \mathcal{N}_B = \mathcal{N}_H$ & $S = L, \mathcal{N}_S = \mathcal{N}_L$\\
			\hline
			\text{If $h_H< h_L$} &$S = H, \mathcal{N}_S = \mathcal{N}_H $  &  $B = L, \mathcal{N}_B = \mathcal{N}_L$\\
			\hline
		\end{tabular}
	\end{table}
	Next, we characterize the types $B$ and $S$ users' equilibrium strategies at the $\overline{\rho}$-SNE and $\underline{\rho}$-SNE in Proposition \ref{Monotonicity}. 
	\begin{prop}[Stage II Equilibrium Strategy]\label{Monotonicity}
		Consider $\rho \in \{\overline{\rho}, \underline{\rho}\}$. The following strategy profile $(\boldsymbol{\lambda}_n^{\rm NE}=$   
		$(\pi_{B}(h_B,h_S,\rho),0),\forall n\in\mathcal{N}_B,\boldsymbol{\lambda}_l^{\rm NE} = (\pi_{S}(h_B,h_S,\rho),0),$   $\forall l\in\mathcal{N}_S)$ constitutes a $\rho$-SNE, where $\pi_{B}(h_B,h_S,\rho)$, $\pi_{S}(h_B,h_S,\rho)$, and the intermediate variables $A_1(h_B,\rho)$ and $A_2(h_B,h_S,\rho)$ are shown in (16)-(\ref{A2}), respectively.
	\end{prop}
	Here we explain the intuition of the $\rho$-SNE, where $\rho\in \{\overline{\rho}, \underline{\rho}\}$. When both $h_B$ and $h_S$ are small (i.e., (16a) and (17a)), users do not generate transactions. When $h_B$ is large but $h_S$ is small (i.e., (16b) and (17b)), only type-$B$ users generate transactions. When both $h_B$ and $h_S$ are large (i.e., (16c) and (17c)), all users generate transactions.
	
	Based on the equilibrium characterized in Proposition \ref{Monotonicity}, we summarize users' equilibria in Theorem \ref{Stage_II_NE2}.
	
	\begin{thm}[Users' Equilibria in Stage II]\label{Stage_II_NE2}\qquad
		\begin{itemize}
			\item If $\Delta(h_B,h_S,\underline{\rho})> \bar{s}\overline{\rho}$, then there exists a $\overline{\rho}$-SNE, where $\Delta(h_B,h_S,\underline{\rho})$ is shown in (\ref{Delta2}).
			\item If $\Delta(h_B,h_S,\underline{\rho})\leqslant \bar{s}\overline{\rho}$, then there exist a $\underline{\rho}$-SNE.
		\end{itemize}
	\end{thm}
	\begin{figure*}
		\begin{equation}\label{Delta2}
		\begin{aligned}
		\Delta(h_B,h_S,\underline{\rho}) = & \max\Big\{h_B-\frac{\gamma}{\mu}-\frac{\gamma\pi_{B}(h_B,h_S,\underline{\rho})[2\mu-N_B\pi_B(h_B,h_S,\underline{\rho})-N_{S}\pi_{S}(h_B,h_S,\underline{\rho})]}{\mu[\mu-N_B\pi_B(h_B,h_S,\underline{\rho})-N_{S}\pi_{S}(h_B,h_S,\underline{\rho})]^2},\\
		&h_S-\frac{\gamma}{\mu}-\frac{\gamma\pi_{S}(h_B,h_S,\underline{\rho})[2\mu-N_B\pi_B(h_B,h_S,\underline{\rho})-N_{S}\pi_{S}(h_B,h_S,\underline{\rho})]}{\mu[\mu-N_B\pi_B(h_B,h_S,\underline{\rho})-N_{S}\pi_{S}(h_B,h_S,\underline{\rho})]^2}\Big\}.
		\end{aligned}
		\end{equation}
		\vspace{-5mm}
	\end{figure*}
	Fig. \ref{Fig_SNE_theorem} illustrates two SNEs in Theorem \ref{Stage_II_NE2} against fee-per-byte choices $(\overline{\rho},\underline{\rho})$, which reflects users' tradeoff between paying low fee and bearing low waiting time. When $\overline{\rho}$ is small, then $\Delta(h_B,h_S,\underline{\rho})> \bar{s}\overline{\rho}$ and a $\overline{\rho}$-SNE exists. In other words, all users choose high fee-per-byte $\overline{\rho}$, because $\overline{\rho}$ is not high enough and hence the consideration of low waiting time dominates the consideration of paying low fee. As $\overline{\rho}$ increases such that $\Delta(h_B,h_S,\underline{\rho})\leqslant \bar{s}\overline{\rho}$, the $\underline{\rho}$-SNE emerges where all users choose low fee-per-byte $\underline{\rho}$. This is because when $\overline{\rho}$ is high, the consideration of paying low fee dominates the consideration of low waiting time. 
	\begin{figure}[h]
		\vspace{-1mm}
		\centerline{\includegraphics[width=3.6cm]{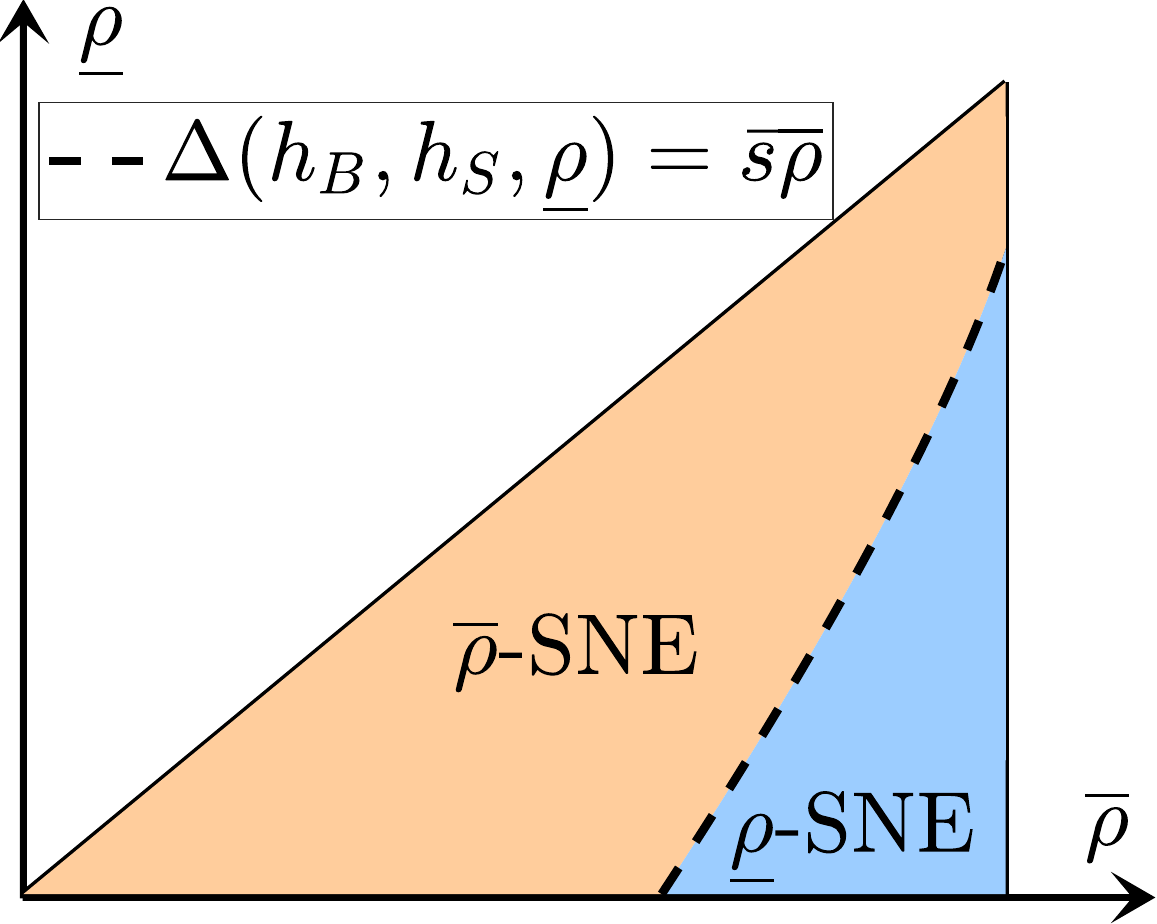}}
		\vspace{-1mm}
		\caption{Two SNEs in Theorem \ref{Stage_II_NE2} VS $\overline{\rho}$ and $\underline{\rho}$.}
		
		\label{Fig_SNE_theorem}
		\vspace{-2mm}
	\end{figure}

	\section{Stage I: Optimal FWT Mechanism of Protocol Designer}\label{Stage_I}
	In this section, we will characterize the protocol designer's optimal FWT mechanism in Stage I. We first formulate the FWT mechanism design as an optimization problem in Section \ref{Stage_I_model}, then we compute its optimal solution in Section \ref{Stage_I_Analysis}.
	\subsection{FWT Mechanism Design of Protocol Designer}\label{Stage_I_model}
	In Stage I, the protocol designer optimizes the FWT mechanism to encourage users to pay sufficient fees while maximizing the social welfare.
	
	\subsubsection{Decision variables} The protocol designer's decision variables are the fee-per-byte choices $\boldsymbol{\rho} = (\overline{\rho},\underline{\rho})$ (with $\overline{\rho}>\underline{\rho}\geqslant 0$) and the waiting tax rate vector $\boldsymbol{P} = (P_{HH}, P_{HL}, P_{LH},P_{LL})$. The fee-per-byte choices encourages users to pay sufficient transaction fees to mitigate the negative externality in transaction selection in Stage III. The waiting tax rate vector let each user internalize the waiting time costs imposed on others, dealing with the negative externality in transaction generation in Stage II.
	
	\subsubsection{Sufficient fee condition}For any user $n$ with a positive transaction generation rate (i.e., $\lambda_{n1}+\lambda_{n2}>0$), the FWT mechanism aims at inducing an average fee-per-byte value that can cover the total storage cost per byte of all miners, that is
	\begin{equation}\label{Sufficient_fee}
	\rho_n^{\text{avg}} = \frac{\lambda_{n1}\overline{\rho}+\lambda_{n2}\underline{\rho}}{\lambda_{n1}+\lambda_{n2}}\geqslant MC_s, \forall n\in\{i\in\mathcal{N}|\lambda_{i1}+\lambda_{i2}>0\}.
	\end{equation}
	\subsubsection{Social welfare} The social welfare equals the sum of users' and miners' time-average payoffs. 
	
	Based on miner $m$'s payoff $v_m^k$ in round $k$ in (\ref{Miner_utility}), the miner $m$'s time-average payoffs as 
	\begin{equation}
	v_{m}(\boldsymbol{\mathcal{X}},\boldsymbol{\rho}) =\lim\limits_{t\rightarrow \infty}\frac{\sum_{k=1}^{\text{Round}(t)} v_m^{k}(\mathcal{X}_m^{k},\boldsymbol{\mathcal{X}}_{-m}^{k},\boldsymbol{\rho})}{t},
	\end{equation}
	where $\boldsymbol{\mathcal{X}} = (\mathcal{X}_m^{k},\forall m,\forall k)$ is the strategy profile of all miners in Stage III and $\text{Round}(t)$ is the number of rounds completed in time interval $[0,t]$.
	
	The social welfare is as follows
	\begin{equation}\label{SW}
	sw(\boldsymbol{\rho},\boldsymbol{P},\boldsymbol{\lambda},\boldsymbol{\mathcal{X}}) =  \sum_{n\in\mathcal{N}}u_n(\boldsymbol{\lambda},\boldsymbol{\rho},\boldsymbol{P})+\sum_{m\in\mathcal{M}}v_{m}(\boldsymbol{\mathcal{X}},\boldsymbol{\rho}).
	\end{equation}
	\subsubsection{FWT mechanism design} We formulate the FWT mechanism design problem in (\ref{SW_max}), which aims at maximizing the social welfare subject to sufficient transaction fee covering the storage cost.
	\begin{equation}\label{SW_max}
	\begin{aligned}
	\max &\hspace{3mm} sw(\boldsymbol{\rho},\boldsymbol{P},\boldsymbol{\lambda},\boldsymbol{\mathcal{X}})\\
	\text{s.t.}\hspace{0.8mm} &\hspace{3mm} \text{(\ref{Sufficient_fee}), }\overline{\rho}>\underline{\rho}\geqslant 0,\\
	\text{var.}\hspace{0.2mm} &\hspace{3mm} \boldsymbol{\rho}=(\overline{\rho},\underline{\rho}),\boldsymbol{P} = (P_{HH}, P_{HL}, P_{LH}, P_{LL}).\text{\footnotemark}
	\end{aligned}
	\end{equation}
	\footnotetext{The waiting tax can be negative, which motivates users to generate transactions by compensating them. This makes the mechanism more flexible.}
	
	It is very challenging to solve Problem (\ref{SW_max}), since it is coupled with the strategies of users in Stage II and miners in Stage III. Nevertheless, we can exploit the special property of the social welfare to derive the optimal solution in closed form.
	
	\subsection{Optimal Solution of FWT Mechanism Design}\label{Stage_I_Analysis}
	In this subsection, we will solve Problem (\ref{SW_max}) and discuss the property of its optimal solution. We will also analyze the effect of the waiting tax rate vector under the optimal FWT mechanism.
	
	\subsubsection{Optimal solution of FWT mechanism design problem}
	The optimal solution to Problem (\ref{SW_max}) is as follows.
	
	\begin{thm}[Optimal Solution of FWT Mechanism Design Problem]\label{Stage_I_NE1}
		The optimal FWT mechanism corresponds to the optimal solution of Problem (\ref{SW_max}) as follows:
		\begin{itemize}[leftmargin=16pt]
			\item If $R_H\leqslant M\bar{s}C_s+\frac{\gamma}{\mu}$, then
			\begin{itemize}
				\item the fee-per-byte choices are $(\overline{\rho}^*,\underline{\rho}^*)  = (MC_s+\frac{\gamma}{\bar{s}\mu},$ $MC_s)$, 
				\item the waiting tax rate vector $(P_{HH}^*, P_{HL}^*, P_{LH}^*,P_{LL}^*)\in\mathbb{R}^{4}$ satisfies the following conditions:
				\begin{equation*}
				\hspace{13.5mm}\begin{cases}
				(N_H-1)P_{HH}^*+N_LP_{HL}^* = 0,\hspace{9mm}\text{\emph{(25a)}}\\
				N_HP_{LH}^*+(N_L-1)P_{LL}^* = 0.\hspace{10.75mm}\text{\emph{(25b)}}\\
				\end{cases}
				\end{equation*}
				\setcounter{equation}{25}
			\end{itemize}
			\item If $R_H> M\bar{s}C_s+\frac{\gamma}{\mu}$, then
			\begin{itemize}
				\item the fee-per-byte choices are $(\overline{\rho}^*,\underline{\rho}^*) =  (\frac{R_H}{\bar{s}} - \frac{\gamma}{\bar{s}\mu},$ $MC_s)$, 
				\item the waiting tax rate vector $(P_{HH}^*, P_{HL}^*, P_{LH}^*,P_{LL}^*)\in\mathbb{R}^{4}$ satisfies the conditions in (26) and the intermediate variables $g_1$ and $g_2$ are shown in (27) and (28), respectively. 
			\end{itemize}
		\end{itemize}
	\end{thm}
	Next we discuss the insights of Theorem \ref{Stage_I_NE1}. 
	If $R_H\leqslant M\bar{s}C_s+\frac{\gamma}{\mu}$, both types of users have low transaction on-chain utilities (i.e., $R_L\leqslant R_H\leqslant M\bar{s}C_s+\frac{\gamma}{\mu}$), which are insufficient to cover a transaction's total storage costs (i.e., $M\bar{s}C_s$) plus waiting time costs (i.e., $\frac{\gamma}{\mu}$). Thus the optimal FWT mechanism prevents both types of users from generating any transactions. For any type-$H$ user (or type-$L$ user, respectively), the sum of his waiting tax payment is 0 as shown in (25a) (or (25b), respectively), due to no transaction generation.

	%
	
	\begin{figure*}
		%
		
		\begin{equation*}\label{Waiting_price_3}
		\hspace{35.5mm}\begin{cases}
		(N_H-1)P_{HH}^*+N_LP_{HL}^* = R_H-\underline{\rho}^*\bar{s}-\dfrac{\gamma[\mu-(N_H-1)g_1-N_Lg_2]}{(\mu-N_Hg_1-N_Lg_2)^2},\hspace{31mm}\text{(26a)}\\
		N_HP_{LH}^*+(N_L-1)P_{LL}^* = R_L-\underline{\rho}^*\bar{s}-\dfrac{\gamma[\mu-N_Hg_1-(N_L-1)g_2]}{(\mu-N_Hg_1-N_Lg_2)^2}.\hspace{33mm}\text{(26b)}\\
		\end{cases}
		\end{equation*}
		\begin{equation*}
		\hspace{4mm}g_1 = \min\Big\{\frac{\mu}{N_H+N_L},\frac{1}{N_H}(\mu-\sqrt{\frac{\gamma\mu}{R_H-M\bar{s}Cs}})\Big\}. \hspace{3mm}\text{(27)}\hspace{11mm}g_2 = \begin{cases}
		0,  &\hspace{-32mm}\text{if }R_L\leqslant M\bar{s}Cs+\frac{\gamma(N_H+N_L)^2}{N_L^2\mu},\\
		\frac{\mu}{N_H+N_L}-\frac{1}{N_L}\sqrt{\frac{\gamma\mu}{R_L-M\bar{s}Cs}}, &\hspace{-2mm}\text{otherwise.}
		\end{cases}\hspace{7mm}\text{(28)}
		\end{equation*}
		\setcounter{equation}{28}
		\vspace{-2mm}
	\end{figure*}

	If $R_H> M\bar{s}C_s+\frac{\gamma}{\mu}$, type-$H$ users have high transaction on-chain utility. Thus the optimal FWT mechanism allows users to generate transactions. The protocol designer sets the low fee-per-byte $\underline{\rho}^* = MC_s$ to guarantee the sufficient fee condition. Since users will generate transactions at SNE, the sum of a type-$H$ user's (or type-$L$ user's, respectively) waiting tax payment is non-zero as shown in (26a) (or (26b), respectively).
	
	\subsubsection{Property of optimal FWT mechanism} To characterize the property of the optimal FWT mechanism, we first establish the benchmark of \emph{unconstrained social optimum}, which is the maximum social welfare that can be achieved without considering the sufficient fee condition (\ref{Sufficient_fee}), i.e., the maximum value of the objective function of Problem (\ref{SW_opt}). 
	\begin{equation}\label{SW_opt}
	\begin{aligned}
	\max &\hspace{3mm} sw(\boldsymbol{\rho},\boldsymbol{P},\boldsymbol{\lambda},\boldsymbol{\mathcal{X}})\\
	\text{s.t.}\hspace{0.8mm} &\hspace{3mm} \overline{\rho}>\underline{\rho}\geqslant 0,\\
	\text{var.}\hspace{0.2mm} &\hspace{3mm} \boldsymbol{\rho}=(\overline{\rho},\underline{\rho}),\boldsymbol{P} = (P_{HH}, P_{HL}, P_{LH}, P_{LL}).
	\end{aligned}
	\end{equation}
	
	Then we characterize the property of the optimal FWT mechanism in Proposition \ref{Optimality}. 
	
	\begin{prop}[Guarantee on Unconstrained Social Optimum]\label{Optimality}
		The maximum social welfare achieved by the optimal FWT mechanism equals the unconstrained social optimum.
	\end{prop}
	Proposition \ref{Optimality} shows through our careful design of the FWT mechanism, imposing the sufficient condition of (\ref{Sufficient_fee}) does not lead to any loss of social welfare. 
	
	\subsubsection{Effect of waiting tax rate vector} Finally, we analyze the effect of the waiting tax rate vector of the optimal FWT mechanism. We define each user $n$'s \emph{total waiting tax rate} $q_n$ based on the waiting tax rate vector defined in Theorem \ref{Stage_I_NE1}, i.e.,
	\begin{equation}
	q_n \triangleq \begin{cases}
	(N_H-1)P_{HH}^*+N_LP_{HL}^*, &\hspace{-0.5mm}\text{if $n\in\mathcal{N}_H$,}\\
	N_HP_{LH}^*+(N_L-1)P_{LL}^*, &\hspace{-0.5mm}\text{if $n\in\mathcal{N}_L$.}\\
	\end{cases}
	\end{equation}
	The total waiting tax rate is the sum of a user's waiting tax payment to all other users for one transaction. We compare the total waiting tax rates of two types of users in Corollary \ref{Waiting_time_price_2}.
	
	\begin{cor}\label{Waiting_time_price_2}
		Under the optimal FWT mechanism, if $R_H> M\bar{s}C_s+\frac{\gamma}{\mu}$, then for any type-H user $\forall n\in\mathcal{N}_H$ and type-L user $\forall l\in\mathcal{N}_L$, we have 
		\begin{itemize}
			\item if $R_H-R_L<\delta$, then $q_n< q_l$, 
			\item if $R_H-R_L\geqslant \delta$, then $q_n\geqslant q_l$, 
		\end{itemize}
		where $\delta = \frac{\gamma\left(g_1-g_2\right)}{\left(\mu-N_{H} g_1-N_{L} g_2\right)^{2}}$ and $g_1$ and $g_2$ are in (27) and (28), respectively.
		%
	\end{cor}

	Corollary \ref{Waiting_time_price_2} leads to an interesting observation. When $R_H-R_L<\delta$, the protocol designer can assign a higher total waiting tax rate for a type-$L$ user $l$ (i.e., $q_n< q_l$), despite such a user $l$ generates fewer transactions and imposes lower waiting time costs on others than a type-$H$ user $n$. The reason is as follows. Without the waiting tax, when $R_L$ is close to $R_H$, a type-$L$ user's transaction generation rate will be close to a type-$H$ user's. However, as $R_H\geqslant R_L$, the optimal FWT mechanism encourages a type-$H$ user to generate much more transactions than a type-$L$ user to maximize the social welfare. Thus, the mechanism assigns a higher total waiting tax rate to force a type-$L$ user to generate fewer transactions to reach the social optimum.

	\section{Performance Evaluations}\label{Evaluation}
	In this section, we evaluate the performance of the optimal FWT mechanism (FWT) by comparing it with the existing blockchain protocol (Existing). We study the impact of various system parameters on the social welfare, fee-per-byte payment, user's payoff, and fairness on both schemes. We further analyze the impact of heterogeneity of miners' storage costs. We summarize the simulation parameters in Table \ref{table2} and we set the blockchain-related parameters based on Ethereum.
	
	\begin{table}[h]\small
		\centering  
		\caption{Blockchain parameters.}
		\label{table2}
		\begin{tabular}{|c|c|}  
			\hline
			Blockchain throughput \cite{No_full_node} & $\mu = 15$ \\
			\hline
			The number of miners \cite{No_full_node} & $M = 10^4$\\
			\hline
			Average transaction size \cite{Avg_size} & $\bar{s} = 150$ bytes \\	
			\hline
			\multirow{2}{*}{Storage cost per byte \cite{Stor_price}} & \multirow{1}{*}{$C_s = 5\times10^{-10}$}\\
			&  \multirow{1}{*}{(USD/byte)}\\
			\hline
			Ratio of transaction on-chain utilities & $R_H:R_L = 2:1$ \\
			\hline
		\end{tabular}
	\end{table}
	
	\subsection{Fee-per-byte and Social Welfare}\label{Num1}
	
	In this subsection, we study how users' parameters (impatience level and transaction on-chain utility) affect both schemes in terms of fee-per-byte and social welfare. For users' parameters, we set the type-$H$ user's transaction on-chain utility as $R_H \in[5\times10^{-4},3\times10^{-3}]$ and the user's impatience level as $\gamma\in[10^{-5},10^{-3}]$. Under such a setting, the daily number of transactions of the existing protocol is between 0.95 to 1.25 millions. This range aligns well with the range of daily number of transactions in Oct. 2020 that is between 0.96 to 1.25 millions.\footnote{\url{https://etherscan.io/chart/tx}}
	
	\subsubsection{Fee-per-byte}
	Fig. 5 (a) and (b) illustrate the impact of impatience level $\gamma$ and transaction on-chain utility $R_H$ on the average fee-per-byte $\rho_n^{\text{avg}}$, respectively.
	\begin{itemize}
		\item Storage cost in this figure corresponds to all miners' total storage cost per byte (i.e., $MC_s$) and serves as a benchmark for the other two curves. Under the optimal FWT mechanism (FWT in the figure), we observe that the average fee-per-byte can cover the total storage cost, satisfying the sufficient fee condition. However, under the existing protocol (Existing in the figure), the sufficient fee condition does not hold when $\gamma\geqslant3.76\times 10^{-3}$. Moreover, we make an interesting observation as follows:
		\begin{observation}\label{Impatience}
			As the impatience level $\gamma$ increases, users pay lower average fee-per-byte $\rho_n^{\text{avg}}$ in the existing protocol.
		\end{observation}
		We explain the reason behind Observation \ref{Impatience} as follows. When the users become more impatient, they generate fewer transactions to reduce waiting costs. Fewer transactions lead to lower incentives to pay high transaction fees and compete for short waiting time.
		\item The correspondences of curves and the legend are the same as Fig. 5(a). Under the optimal FWT mechanism, the system always satisfies the sufficient fee condition. Under the existing protocol, users increase the average fee-per-byte with $R_H$ and the sufficient fee condition only holds when $R_H\geqslant1.6\times 10^{-3}$. Here we explain the reason for the fee-per-byte increase. As the user's on-chain utility $R_H$ increases, users have higher incentives to pay high transaction fees and compete for short waiting time. 
	\end{itemize}
	
	\begin{figure}[t]
		\begin{minipage}[t]{0.48\textwidth}
			\centering
			\subfigure[Users' average fee-per-byte vs. impatience level $\gamma$ for $R_H = 1.8\times10^{-3}$.]{
				\includegraphics[width=0.445\linewidth]{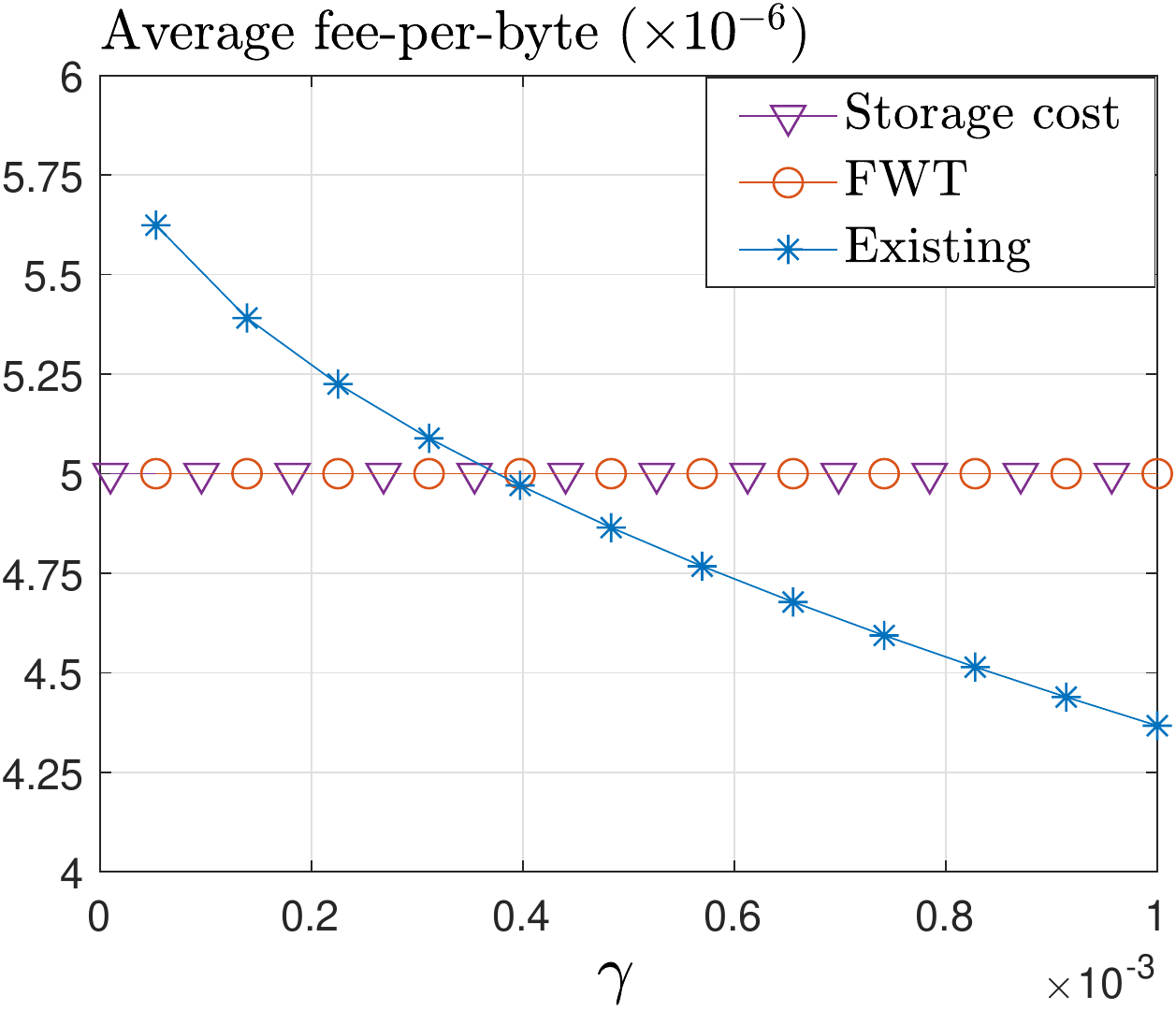}}
			\subfigure[Users' average fee-per-byte vs. type-$H$ users' transaction on-chain utility $R_H$ for $\gamma = 5\times10^{-5}$.]{
				\includegraphics[width=0.435\linewidth]{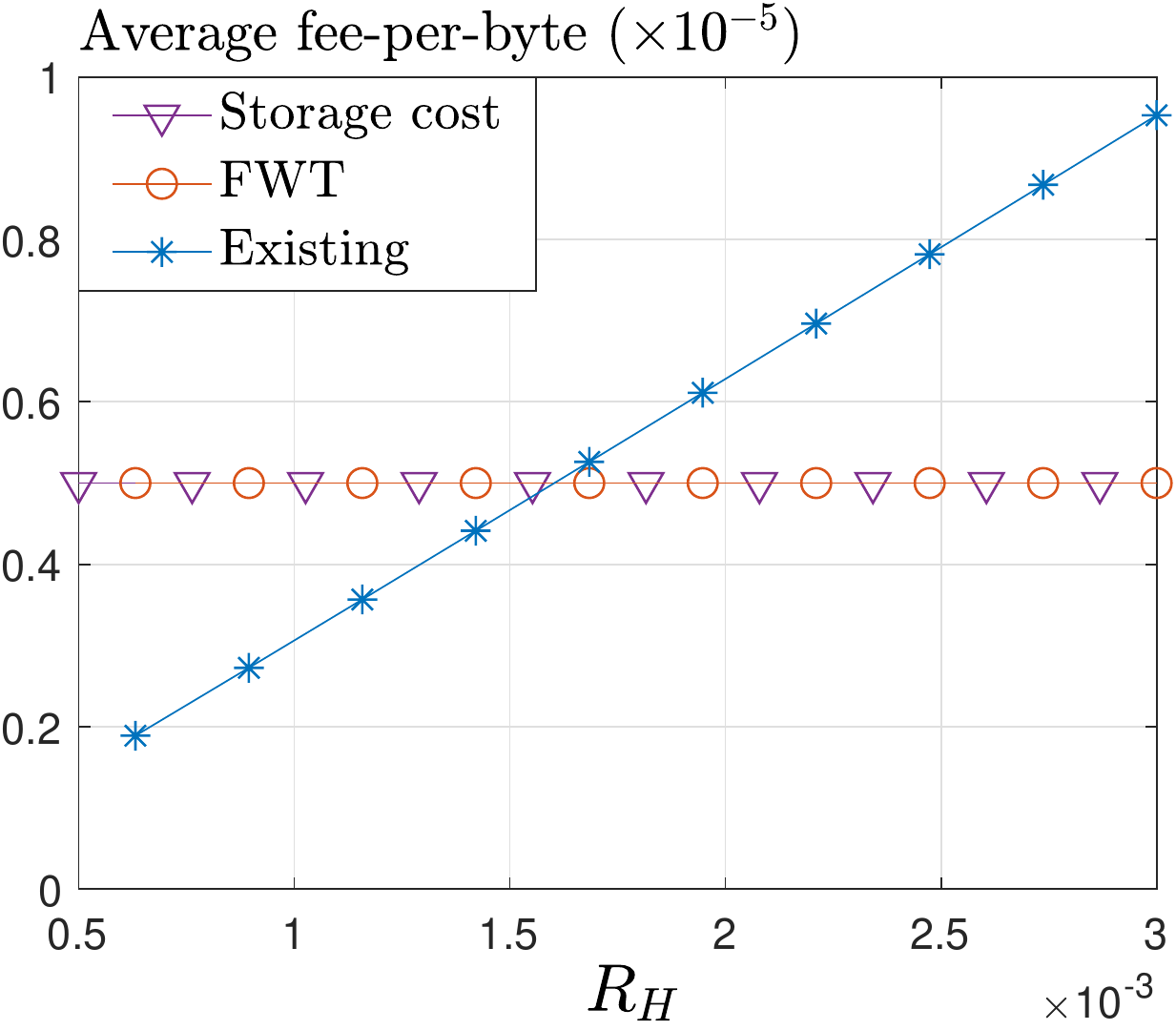}}
			\vspace{-2mm}
			\caption{Impact of users' parameters on average fee-per-byte $\rho_n^{\text{avg}}$.}
		\end{minipage}
		\begin{minipage}[t]{0.48\textwidth}
			\centering
			\subfigure[Social welfare and improvement vs. impatience level $\gamma$ for $R_H = 1.8\times10^{-3}$.]{
				\includegraphics[width=0.46\linewidth]{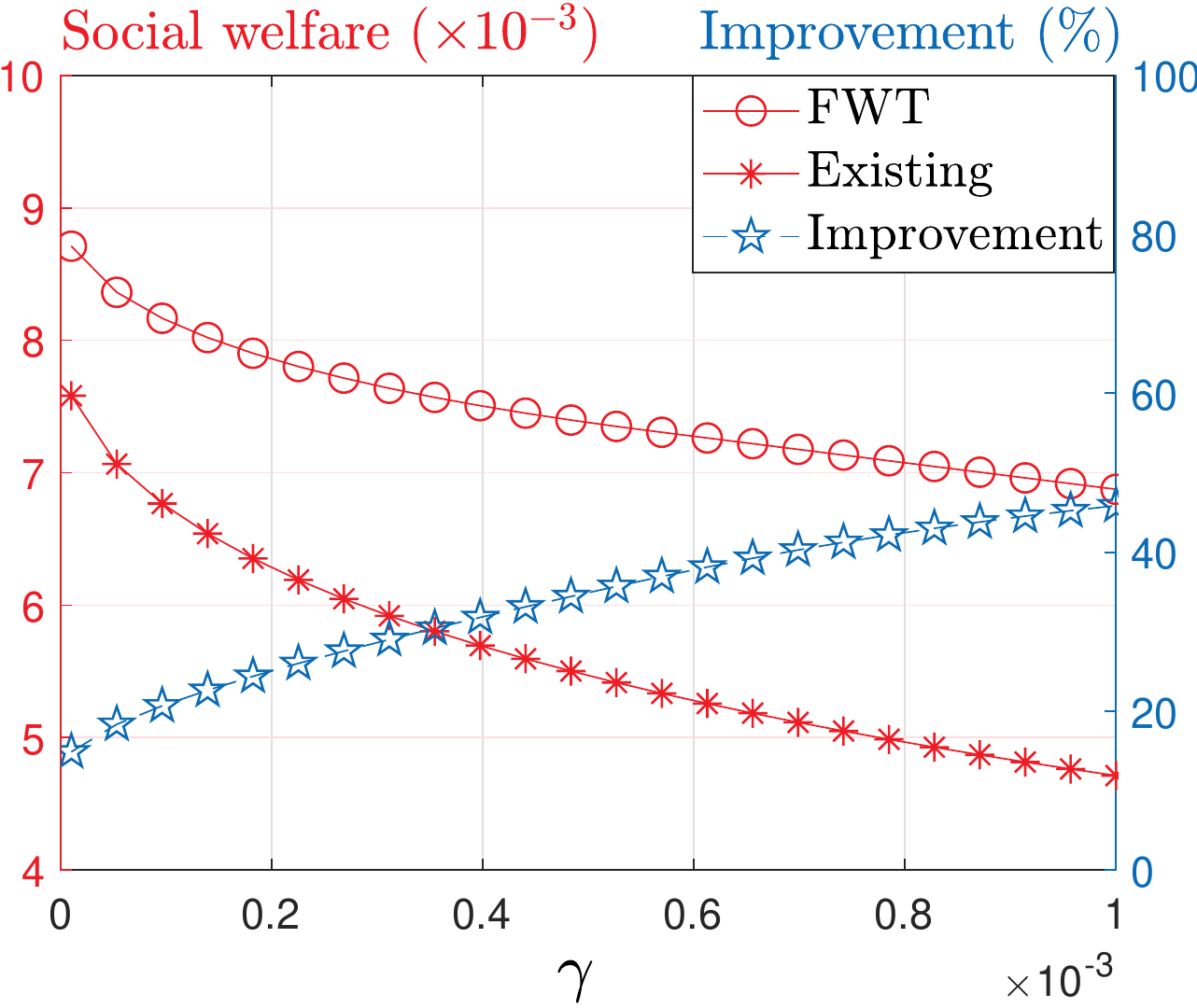}}
			\subfigure[Social welfare and improvement vs. type-$H$ users' on-chain utility $R_H$ for $\gamma = 5\times10^{-5}$.]{
				\includegraphics[width=0.46\linewidth]{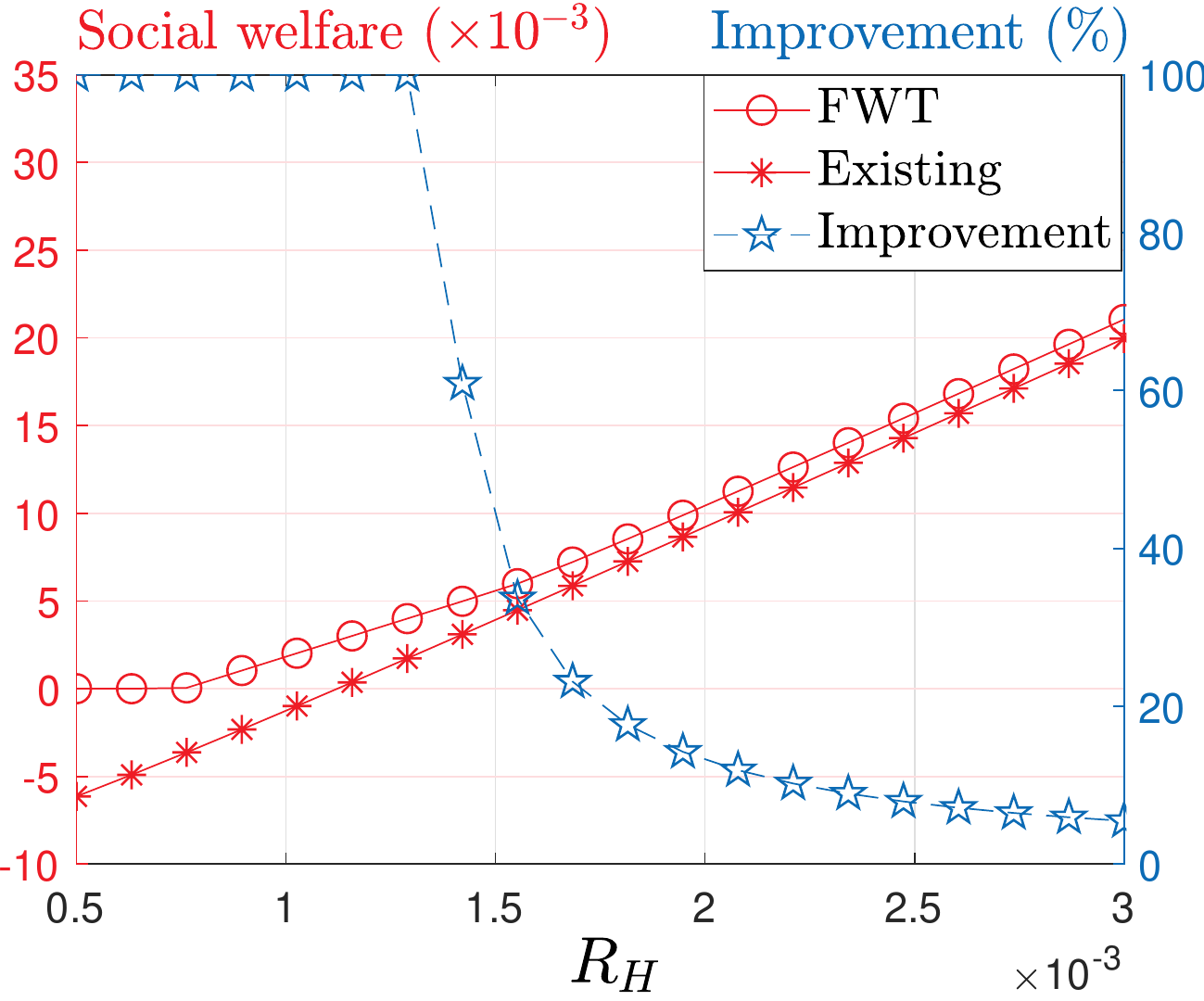}}
			\vspace{-2mm}
			\caption{Impact of users' parameters on social welfare $sw$.}
		\end{minipage}
		\vspace{-2mm}
	\end{figure}
	

	\subsubsection{Social welfare}\label{Num_gamma}
	Fig. 6 (a) and (b) illustrate the impact of impatience level $\gamma$ and transaction on-chain utility $R_H$ on the social welfare $sw$, respectively.
	\begin{itemize}
		\item Fig. 6(a): On the left axis, two red curves plot the social welfares of the optimal FWT mechanism (FWT in the figure) and the existing protocol (Existing in the figure). We notice that the social welfares of both schemes decrease in $\gamma$, due to the increased waiting time cost with the increasing impatience level $\gamma$. On the right axis, the blue curve marked in stars plots the optimal FWT mechanism's social welfare improvement over the existing protocol (Improvement in the figure). We observe that the social welfare improvement increases in $\gamma$ with an average value of 33.73\%. The reason for such an improvement is that the optimal FWT mechanism addresses the negative externality in transaction generation and reduces the transaction waiting time. As the waiting time cost increases with $\gamma$, the consideration of the negative externality brings more social welfare improvement. 
		\item Fig. 6(b): The correspondences of curves and the axes are similar as Fig. 6(a). On the left axis, we observe that the social welfares of both schemes increase in $R_H$, due to the increased on-chain utility. On the right axis, the social welfare improvement decreases in $R_H$ with an average value of 45.68\%. The reason for such a decrease in improvement is as follows. In the existing protocol, the average fee-per-byte increases with $R_H$ (i.e., Fig 5(b)), preventing users from generating too many transactions and causing excessive waiting time costs on others.
	\end{itemize}
	Based on Figs. 5 and 6, we have the following observation:
	\begin{observation}\label{Obs_improve}
		The optimal FWT mechanism achieves average social welfare of 33.73\% or more compared with the existing blockchain protocol while guarantees that users pay sufficient fees.
	\end{observation}
	Observation \ref{Obs_improve} demonstrates that the optimal FWT mechanism dominates existing protocol in both social welfare and sufficient fees for covering storage costs.

	\subsection{User's Payoff and Fairness}
	

	\begin{figure}[htbp]
		\vspace{-1mm}
		\begin{minipage}[t]{0.23\textwidth}
			\centering
			\label{Payoff_RH}
			\includegraphics[width=0.93\linewidth]{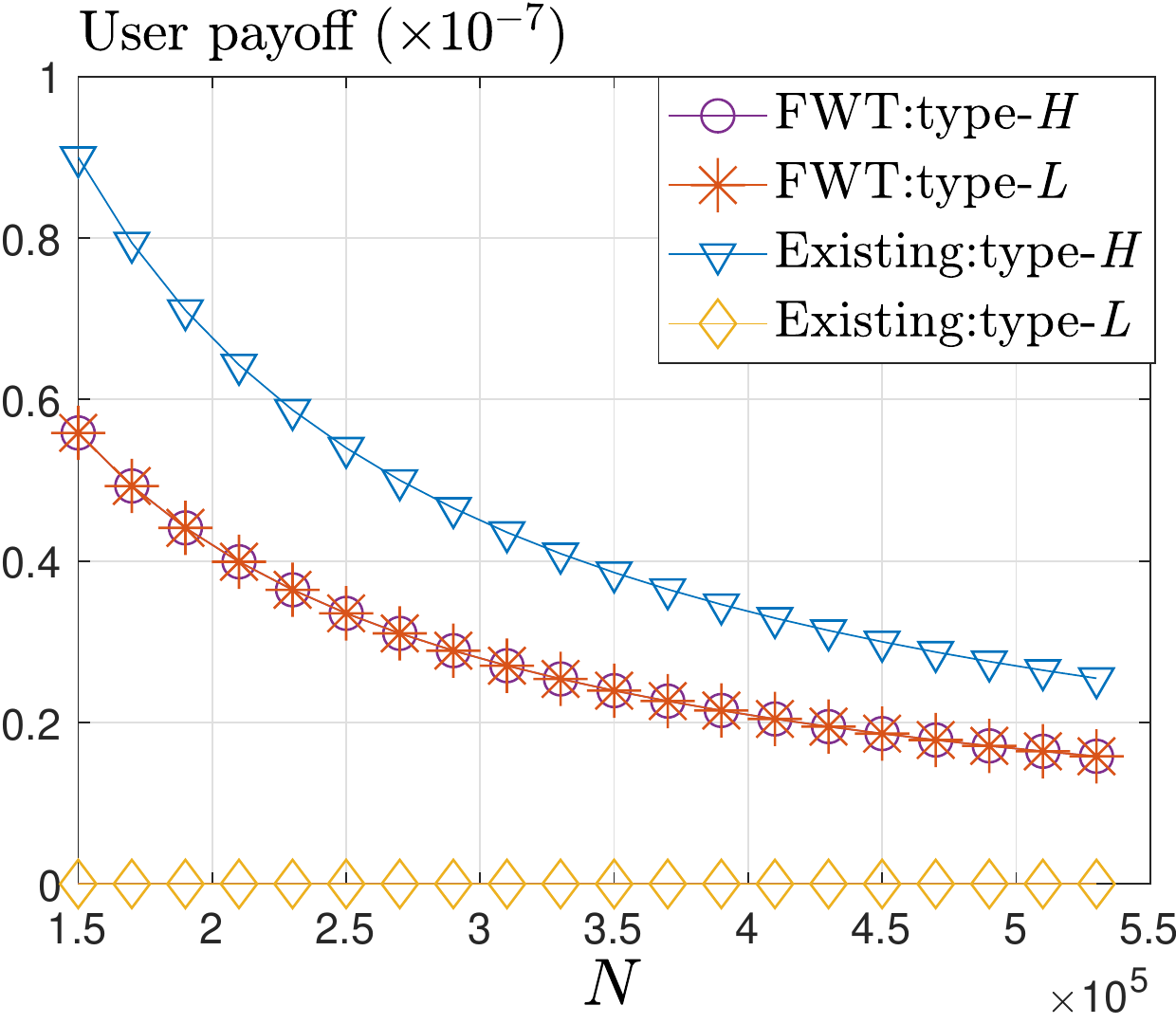}
			\vspace{-1mm}
			\caption{Users' payoffs vs. numbers of users $N$.}
		\end{minipage}\quad
		\begin{minipage}[t]{0.23\textwidth}
			\centering
			\label{Fair_RH}
			\includegraphics[width=0.94\linewidth]{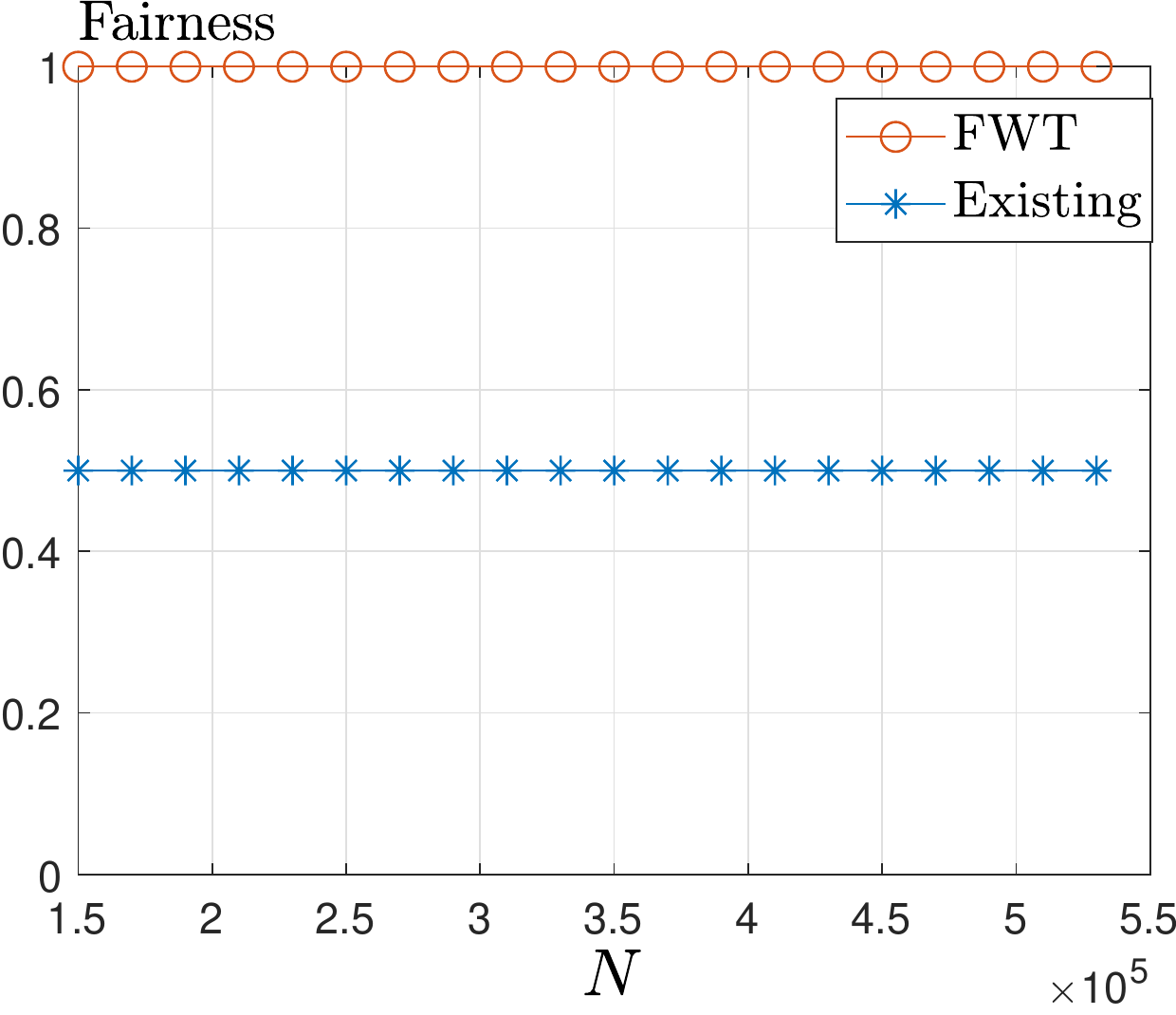}
			\vspace{-1mm}
			\caption{Jain’s fairness index vs. numbers of users $N$.}
		\end{minipage}
		\vspace{-2mm}
	\end{figure}
	
	In this subsection, we analyze the impact of the numbers of users $N$ on the user's payoff and fairness. The range of $N$ is set based on minimum and maximum of daily active users of Ethereum in 2020, i.e., $N\in[153\text{k},537\text{k}]$. Users' parameters are set as $R_H = 1.8\times10^{-3}$ and $\gamma = 5\times10^{-5}$. 
	
	In Fig. 7, we plot the type-$H$ and type-$L$ users' payoffs under two schemes against $N$. Users' payoffs decrease in $N$ under both schemes, as more users compete for short waiting time and each user will generate fewer transactions. Moreover, the payoff difference between a type-$H$ user and a type-$L$ user is lower under the optimal FWT mechanism than under the existing protocol. This is because type-$H$ users compensate type-$L$ users by the waiting tax.
	
	In Fig. 8, we study the users' payoffs in terms of Jain's fairness index \cite{Fairness} (defined as $\left(\sum\nolimits_{n \in \mathcal{N}} u_{n}\right)^{2} /\left(N \sum_{n \in \mathcal{N}} u_n^{2}\right)$) again $N$. The index measures the fairness level of all users' payoffs. We makes the following observation: 
	\begin{observation}\label{Fairness}
		The optimal FWT mechanism achieves the maximum fairness index of 1, which is higher than the existing protocol.
	\end{observation} 
	Here we explain the reason for achieving the maximum fairness index. Under the optimal FWT mechanism, each user's payment on the waiting tax perfectly compensates others for the waiting time costs he imposes. The tax can balance users' payoffs despite users have heterogeneous on-chain utilities.
	
	\subsection{Impact of Heterogeneous Storage Costs}\label{Num2}
	\begin{figure}[htbp]
		\label{User_uncertain}
		\centering
		\subfigure[User's average fee-per-byte vs. storage costs ratio $C_{s,H}/C_{s,L}$.]{
			\includegraphics[width=0.434\linewidth]{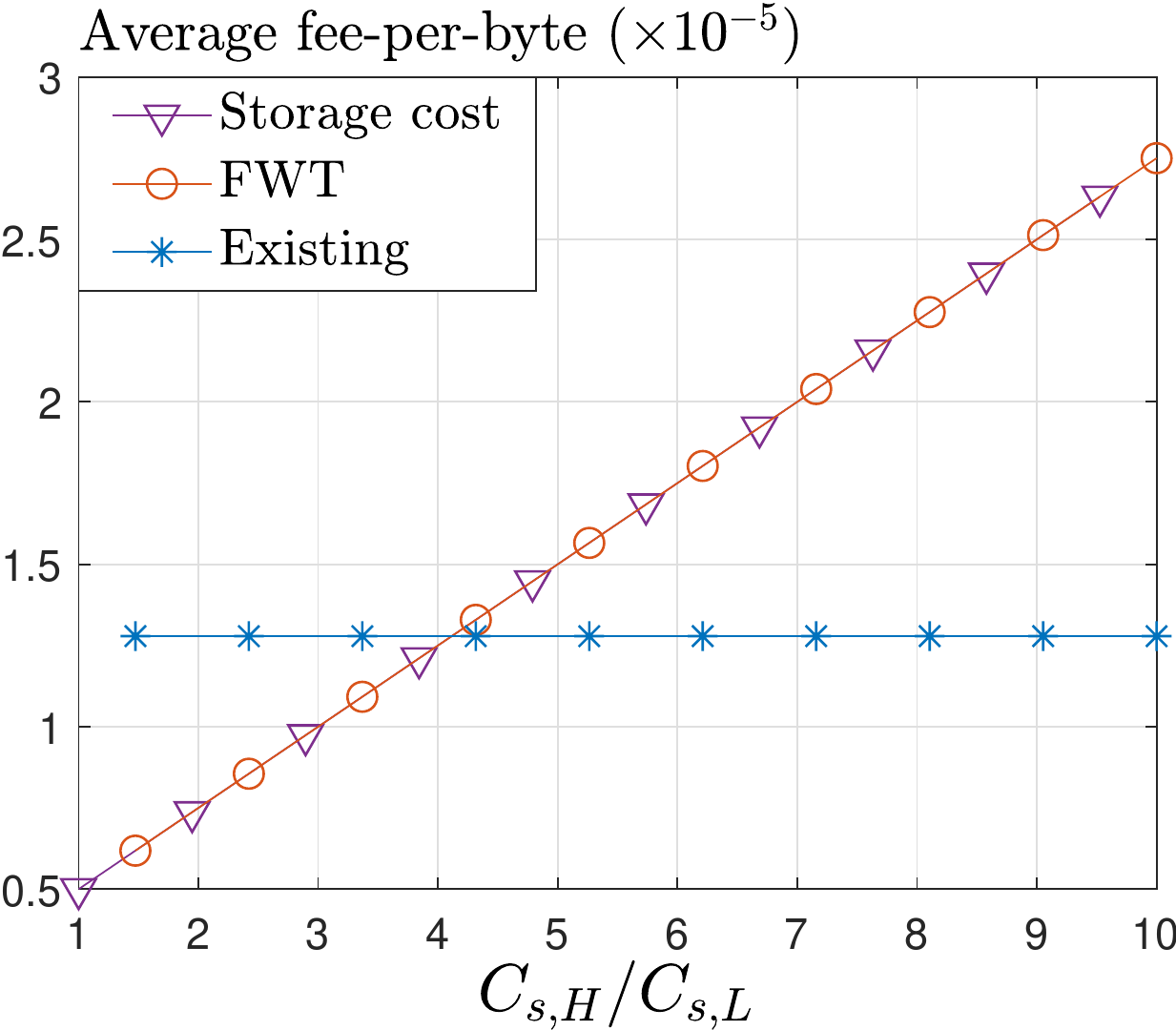}}
		\subfigure[Social welfare and improvement vs. storage costs ratio $C_{s,H}/C_{s,L}$.]{
			\includegraphics[width=0.45\linewidth]{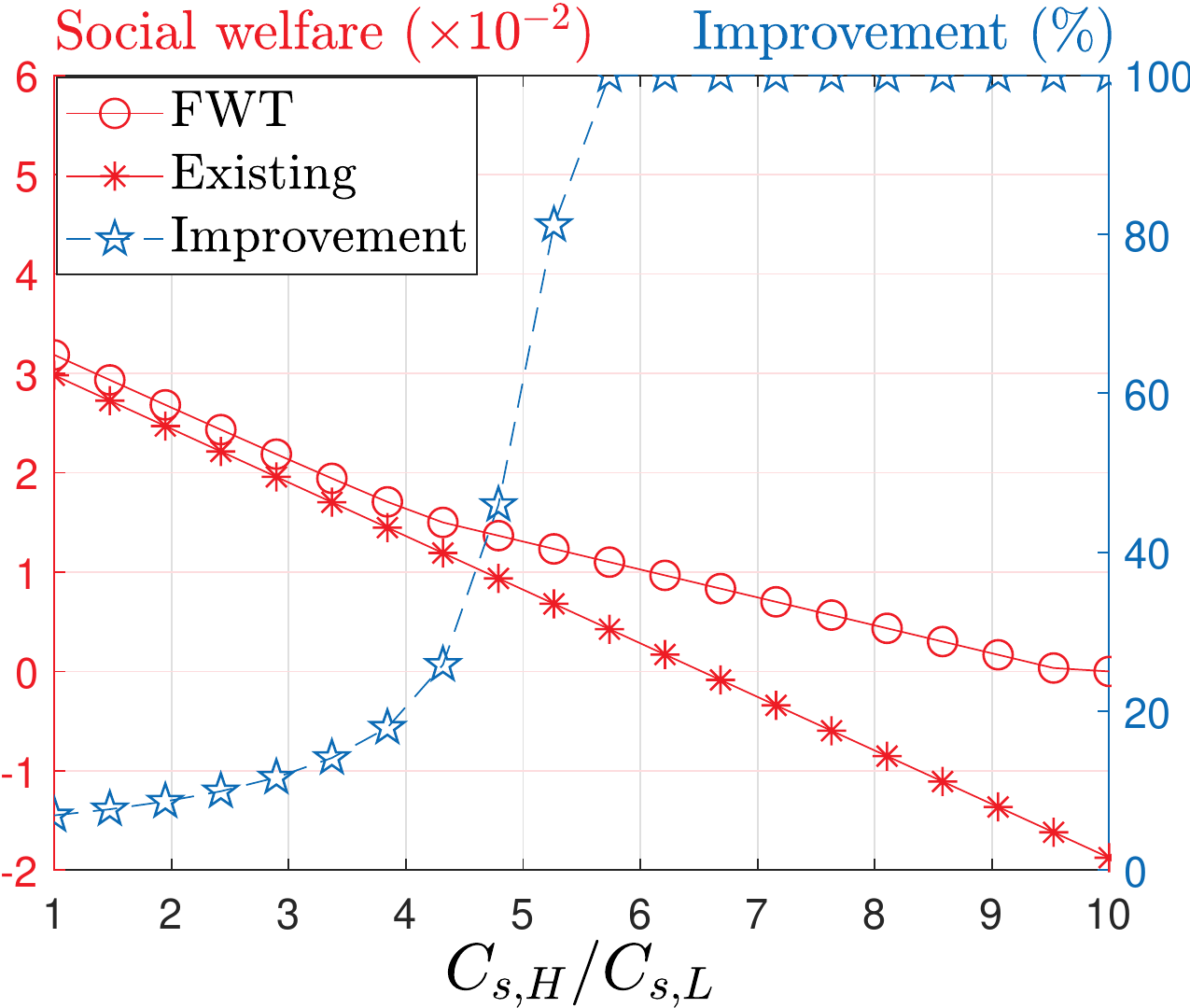}}
		\vspace{-2mm}
		\caption{Impact of heterogeneity of storage costs.}
		\vspace{-2mm}
	\end{figure}
	So far, we have studied the blockchain mechanism design under homogeneous-storage-cost miners. However, miners' storage costs can be heterogeneous (e.g., Ethereum offers different data storage modes consuming different amounts of storage), which will affect the performance of the mechanism. In this subsection, we study the impact of such heterogeneity on our FWT mechanism and the existing blockchain protocol.
	
	We conduct the numerical analysis to analyze the impact of heterogeneous storage costs. According to the two storage modes in Ethereum, we assume that half of the miners have high storage costs while the other half have low storage costs, denoted by $C_{s,H}$ and $C_{s,L}$ per byte, respectively. We analyze the storage costs ratio $C_{s,H}/C_{s,L}$ within the range of $ [1,10]$ based on the actual settings of two Ethereum storage modes.\footnote{Ethereum mainly offers full node synchronization mode and archival mode. On Jan. 2021, full node synchronization mode requires 610 GB of disk space while an archival node requires 6.0 TB.} We choose the low storage cost per byte based on the price of SSD disk per byte, i.e., $C_{s,L} = 5\times10^{-10}$ (USD/byte) \cite{Stor_price}.
	Users' parameters are set as $R_H = 4\times10^{-3}$ and $\gamma = 5\times10^{-5}$.
	
	For the FWT mechanism design problem (\ref{SW_max}), we need to reformulate it considering different storage costs, i.e.,
	\vspace{-2mm}
	\begin{equation}\label{Mech2}
	\begin{aligned}
	\max &\hspace{1.1mm} sw(\boldsymbol{\rho},\boldsymbol{P},\boldsymbol{\lambda},\boldsymbol{\mathcal{X}})\\
	\text{s.t.}\hspace{0.8mm} &\hspace{1.1mm} \rho_n^{\text{avg}} \geqslant \frac{M(C_{s,L}+C_{s,H})}{2},\forall n\in\{i\in\mathcal{N}|\lambda_{i1}+\lambda_{i2}>0\},\\
	\text{    }\hspace{0.8mm} &\hspace{1.1mm} \overline{\rho}>\underline{\rho}\geqslant 0,\\
	\text{var.}\hspace{0.2mm} &\hspace{1.1mm} \boldsymbol{\rho}=(\overline{\rho},\underline{\rho}), \boldsymbol{P} = (P_{HH}, P_{HL}, P_{LH}, P_{LL}).
	\end{aligned}
	\end{equation}
	The only difference between Problems (\ref{Mech2}) and (\ref{SW_max}) is the constraint of sufficient fee condition, due to the difference of storage costs.
	
	To simplify Problem (\ref{Mech2}), we only consider the feasible region $\overline{\rho}>\underline{\rho}\geqslant \frac{M(C_{s,L}+C_{s,H})}{2}$, such that any point within the region always satisfies the constraint. Within the region, we prove that each miner's equilibrium strategy is the same as Theorem \ref{Stage_III_NE}, since the fee-per-byte choices are sufficiently high and no miners reject any transaction. Thus, users' equilibrium within the region is also the same as Theorem \ref{Stage_II_NE2}. Finally, we can solve Problem (\ref{Mech2}) just like we solve Problem (\ref{SW_max}). The details are in \cite{Online_Appendix}.
	
	In Fig. 9, we illustrate the impact of storage costs' heterogeneity on the average fee-per-byte $\rho_n^{\text{avg}}$ and social welfare $sw$. 
	
	\begin{itemize}
		\item Fig. 9(a): Storage cost in this figure shows all miners' total storage cost per byte (i.e., $\frac{M(C_{s,L}+C_{s,H})}{2}$) and serves as a benchmark for the other two curves. Under the optimal FWT mechanism, the average fee-per-byte can cover the total storage cost, satisfying the sufficient fee condition. However, in the existing protocol, the average fee-per-byte does not change with $C_{s,H}/C_{s,L}$ and the sufficient fee condition does not hold when $C_{s,H}/C_{s,L}\geqslant 4.12$. The reason for the unchanged fee-per-byte is as follows. We consider low-fee transactions that low-storage-cost miners admit. If high-storage-cost miners do not admit those transactions, they may bear the storage costs and get no fee. Thus, they prefer to admit, such that they bear the storage costs but get some fees. Then users do not need to pay a higher fee-per-byte as $C_{s,H}/C_{s,L}$ increases.
		\item Fig. 9(b): The correspondences of curves and the axes are similar as Fig. 6(a). On the left axis, the social welfares of both schemes decrease in the storage costs ratio $C_{s,H}/C_{s,L}$, due to the increased storage costs. On the right axis, the social welfare improvement of the optimal FWT mechanism over the existing protocol increases in $C_{s,H}/C_{s,L}$ with an average value of 61.49\%. The improvement is due to the optimal FWT mechanism considers miners' negative externality in transaction selection.
	\end{itemize}
	
	Based on Fig. 9, we have the following observation:
	
	\begin{observation}\label{Obs_uncertain2}
		Under heterogeneous-storage-cost miners, the optimal FWT mechanism still outperforms the existing protocol in both social welfare and sufficient fees for covering storage costs.
	\end{observation}
	Observation \ref{Obs_uncertain2} demonstrates the effectiveness of the optimal FWT mechanism under heterogeneous-storage-cost miners.
	
	\section{Conclusion}\label{Conclusion}
	In this paper, we proposed an FWT mechanism to mitigate the issue of insufficient storage fee in blockchain. We noticed two types of negative externalities in the system: a miner's transaction selection imposes storage costs on other miners and a user's transaction generation imposes waiting time costs on other users. Motivated by the negative externalities, the FWT mechanism offers fee choices to users and imposes waiting tax on them. We modeled the interactions among the protocol designer, users, and miners as a three-stage Stackelberg game. We derived the subgame perfect equilibrium of the game in closed-form. Based on the equilibrium, we found that miners neglecting the negative externality in transaction selection cause the insufficient fee issue in the existing blockchain. We showed that the optimal FWT mechanism achieves the unconstrained social optimum and guarantees that users pay sufficient transaction fees for storage costs. Surprisingly, we further found that users who impose lower waiting time costs on other users may pay a higher waiting tax under the optimal FWT mechanism, as the mechanism encourages other users to generate more transactions to maximize the social welfare. Ethereum-based numerical results showed that the optimal FWT mechanism guarantees sufficient transaction fees and achieves an average social welfare improvement of 33.73\% or more over the existing protocol. Furthermore, the optimal FWT mechanism achieves the maximum fairness index, and performs well even under heterogeneous-storage-cost miners.
	
	In future work, we will extend our analysis to the more practical case where the number of miners and users in the system may dynamically change over time. 
	\vspace{-5pt}

	\bibliographystyle{IEEEtran}
	\bibliography{Eth_storage} 

\begin{thebibliography}{10}
\providecommand{\url}[1]{#1}
\csname url@samestyle\endcsname
\providecommand{\newblock}{\relax}
\providecommand{\bibinfo}[2]{#2}
\providecommand{\BIBentrySTDinterwordspacing}{\spaceskip=0pt\relax}
\providecommand{\BIBentryALTinterwordstretchfactor}{4}
\providecommand{\BIBentryALTinterwordspacing}{\spaceskip=\fontdimen2\font plus
\BIBentryALTinterwordstretchfactor\fontdimen3\font minus
  \fontdimen4\font\relax}
\providecommand{\BIBforeignlanguage}[2]{{%
\expandafter\ifx\csname l@#1\endcsname\relax
\typeout{** WARNING: IEEEtran.bst: No hyphenation pattern has been}%
\typeout{** loaded for the language `#1'. Using the pattern for}%
\typeout{** the default language instead.}%
\else
\language=\csname l@#1\endcsname
\fi
#2}}
\providecommand{\BIBdecl}{\relax}
\BIBdecl

\bibitem{liueconomics}
Y.~Liu, Z.~Fang, M.~H. Cheung, W.~Cai, and J.~Huang, ``Economics of blockchain
  storage,'' in \emph{IEEE International Conference on Communications}, 2020,
  pp. 1--6.

\bibitem{Archive_node}
\url{https://medium.com/quiknode/welcoming-the-newest-member-of-the}
  \url{-quiknode-family-the-eth-archive-node-ac66201e0793}.

\bibitem{Blockchain_survey}
W.~Wang, D.~T. Hoang, P.~Hu, Z.~Xiong, D.~Niyato, P.~Wang, Y.~Wen, and D.~I.
  Kim, ``A survey on consensus mechanisms and mining strategy management in
  blockchain networks,'' \emph{IEEE Access}, vol.~7, pp. 22\,328--22\,370,
  2019.

\bibitem{Ethsize}
\url{https://etherscan.io/chartsync/chainarchive}.

\bibitem{Eth_fee}
\url{https://studio.glassnode.com/compare?a=ETH&c=usd&e=&m=fees.VolumeSum&mAvg=0&mMedian=0&mScl=lin&miner=&resolution=1month&s=1440896401&sameAxis=true&u=1601510400&zoom=}.

\bibitem{No_full_node}
\url{https://www.ethernodes.org/history}.

\bibitem{Bitcoin_white_paper}
S.~Nakamoto, ``Bitcoin: A peer-to-peer electronic cash system,''
  https://bitcoin.org/bitcoin.pdf, 2008.

\bibitem{Nodedecline}
\url{https://thebitcoin.pub/t/ethereum-essentials-node-nuances/52785}.

\bibitem{BitcoinWiki_Weaknesses}
``Bitcoinwiki. weaknesses,'' \url{https://en.bitcoin.it/wiki/Weaknesses}.

\bibitem{cai2018decentralized}
W.~Cai, Z.~Wang, J.~B. Ernst, Z.~Hong, C.~Feng, and V.~C.~M. Leung,
  ``Decentralized applications: The blockchain-empowered software system,''
  \emph{IEEE Access}, vol.~6, pp. 53\,019--53\,033, 2018.

\bibitem{Storage_discuss}
\url{https://ethereum-magicians.org/search?q=state\%20rent}.

\bibitem{Monopoly_blockchain}
G.~Huberman, J.~Leshno, and C.~C. Moallemi, ``Monopoly without a monopolist: An
  economic analysis of the bitcoin payment system,'' \emph{Bank of Finland
  Research Discussion Paper}, no.~27, 2017.

\bibitem{Mining_to_market}
D.~Easley, M.~O'Hara, and S.~Basu, ``From mining to markets: The evolution of
  bitcoin transaction fees,'' \emph{Journal of Financial Economics}, vol. 134,
  no.~1, pp. 91--109, 2019.

\bibitem{Transaction_Queuing}
J.~Li, Y.~Yuan, S.~Wang, and F.-Y. Wang, ``Transaction queuing game in bitcoin
  blockchain,'' in \emph{IEEE Intelligent Vehicles Symposium}, 2018, pp.
  114--119.

\bibitem{Block_size}
P.~R. Rizun, ``A transaction fee market exists without a block size limit,''
  \emph{Block Size Limit Debate Working Paper}, 2015.

\bibitem{Feewoblocksize2}
R.~Zhang and B.~Preneel, ``On the necessity of a prescribed block validity
  consensus: Analyzing bitcoin unlimited mining protocol,'' in \emph{ACM
  International Conference on Emerging Networking Experiments and
  Technologies}, 2017, pp. 108--119.

\bibitem{EIP_1559}
\url{https://github.com/ethereum/EIPs/blob/master/EIPS/eip-1559.md}.

\bibitem{Fee_design_ABC}
Q.~Hu, Y.~Nigam, Z.~Wang, Y.~Wang, and Y.~Xiao, ``A correlated equilibrium
  based transaction pricing mechanism in blockchain,'' in \emph{IEEE
  International Conference on Blockchain and Cryptocurrency}, 2020, pp. 1--7.

\bibitem{Fee_design_Auction}
Z.~Ai, Y.~Liu, and X.~Wang, ``Abc: An auction-based blockchain
  consensus-incentive mechanism,'' in \emph{IEEE International Conference on
  Parallel and Distributed Systems}, 2020, pp. 609--616.

\bibitem{Fee_design_fee_market}
S.~Basu, D.~Easley, M.~O'Hara, and E.~Sirer, ``Towards a functional fee market
  for cryptocurrencies,'' \emph{Available at SSRN 3318327}, 2019.

\bibitem{Fee_design_re_fee_market}
R.~Lavi, O.~Sattath, and A.~Zohar, ``Redesigning bitcoin's fee market,'' in
  \emph{ACM The World Wide Web Conference}, 2019, pp. 2950--2956.

\bibitem{Bitcoinbook}
A.~Narayanan, J.~Bonneau, E.~Felten, A.~Miller, and S.~Goldfeder, \emph{Bitcoin
  and cryptocurrency technologies: a comprehensive introduction}.\hskip 1em
  plus 0.5em minus 0.4em\relax Princeton University Press, 2016.

\bibitem{Fee_per_byte}
https://metamug.com/article/security/bitcoin-transaction-fee-satoshi-per-byte.

\bibitem{Blockchain_SW1}
Y.~Jiao, P.~Wang, D.~Niyato, and K.~Suankaewmanee, ``Auction mechanisms in
  cloud/fog computing resource allocation for public blockchain networks,''
  \emph{IEEE Transactions on Parallel and Distributed Systems}, vol.~30, no.~9,
  pp. 1975--1989, 2019.

\bibitem{Blockchain_SW2}
C.~Chen, J.~Wu, H.~Lin, W.~Chen, and Z.~Zheng, ``A secure and efficient
  blockchain-based data trading approach for internet of vehicles,'' \emph{IEEE
  Transactions on Vehicular Technology}, vol.~68, no.~9, pp. 9110--9121, 2019.

\bibitem{Eth_Top_Fee}
``Top 50 gas spenders,'' \url{https://etherscan.io/gastracker#gassender}.

\bibitem{BlockGeneration2018}
R.~Bowden, H.~P. Keeler, A.~E. Krzesinski, and P.~G. Taylor, ``Block arrivals
  in the bitcoin blockchain,'' \emph{arXiv preprint arXiv:1801.07447}, 2018.

\bibitem{Gap_Game}
I.~Tsabary and I.~Eyal, ``The gap game,'' in \emph{ACM SIGSAC Conference on
  Computer and Communications Security}, 2018, pp. 713--728.

\bibitem{Block_propagation_time}
C.~Decker and R.~Wattenhofer, ``Information propagation in the bitcoin
  network,'' in \emph{IEEE International Conference on Peer-to-Peer Computing},
  2013, pp. 1--10.

\bibitem{Myopic_mining}
R.~Singh, A.~D. Dwivedi, G.~Srivastava, A.~Wiszniewska-Matyszkiel, and
  X.~Cheng, ``A game theoretic analysis of resource mining in blockchain,''
  \emph{Cluster Computing}, vol.~23, no.~3, pp. 2035--2046, 2020.

\bibitem{Fruitchains}
R.~Pass and E.~Shi, ``Fruitchains: A fair blockchain,'' in \emph{ACM Symposium
  on Principles of Distributed Computing}, 2017, pp. 315--324.

\bibitem{IOTA}
\url{https://www.iota.org/}.

\bibitem{Fee_per_byte1}
C.~Wang, X.~Chu, and Y.~Qin, ``Measurement and analysis of the bitcoin
  networks: A view from mining pools,'' in \emph{IEEE International Conference
  on Big Data Computing and Communications}, 2020, pp. 180--188.

\bibitem{Fee_recommendation1}
\url{https://bitcoiner.live/?tab=info}.

\bibitem{Fee_recommendation2}
\url{https://www.buybitcoinworldwide.com/fee-calculator/}.

\bibitem{Online_Appendix}
``Online appendix,''
  \url{https://www.dropbox.com/s/otoi915dms9g6k6/Liuyunshu_OnlineAppendix.pdf?dl=0}.

\bibitem{queue_book}
D.~Gross, \emph{Fundamentals of queueing theory}.\hskip 1em plus 0.5em minus
  0.4em\relax John Wiley \& Sons, 2008.

\bibitem{huang2021eliciting}
C.~Huang, H.~Yu, J.~Huang, and R.~A. Berry, ``Crowdsourcing with heterogeneous
  workers in social networks,'' in \emph{IEEE Global Communications
  Conference}, 2019, pp. 1--6.

\bibitem{Avg_size}
https://ethereum.stackexchange.com/questions/30175/what-is-the-size-
  bytes-of-a-simple-ethereum-transaction-versus-a-bitcoin-trans.

\bibitem{Stor_price}
\url{https://www.amazon.com/ssd/s?k=ssd}.

\bibitem{Fairness}
R.~K. Jain, D.-M.~W. Chiu, W.~R. Hawe \emph{et~al.}, ``A quantitative measure
  of fairness and discrimination,'' \emph{Technical Report}, Digital Equipment
  Corporation, Hudson, MA, 1984.

\end{thebibliography}
	
\end{document}